\documentclass[trackchanges,twocolumn]{aastex}

\let\tablenum\relax

\usepackage{natbib,epsfig,siunitx,url,graphicx,amsmath,footnote,float,xcolor,cases}
\begin{document}

\submitted{Accepted for publication in Icarus}

\title{The early instability scenario: Mars' mass explained by Jupiter's orbit}

\author{Matthew S. Clement\altaffilmark{1}, Nathan A. Kaib\altaffilmark{2}, Sean N. Raymond\altaffilmark{3} \& John E. Chambers\altaffilmark{1}}

\altaffiltext{1}{Earth and Planets Laboratory, Carnegie Institution for Science, 5241 Broad Branch Road, NW, Washington, DC 20015, USA}
\altaffiltext{2}{HL Dodge Department of Physics Astronomy, University of Oklahoma, Norman, OK 73019, USA}
\altaffiltext{3}{Laboratoire d'Astrophysique de Bordeaux, Univ. Bordeaux, CNRS, B18N, all{\'e} Geoffroy Saint-Hilaire, 33615 Pessac, France}
\altaffiltext{*}{corresponding author email: mclement@carnegiescience.edu}

\begin{abstract}

The formation of the solar system's giant planets predated the ultimate epoch of massive impacts that concluded the process of terrestrial planet formation.  Following their formation, the giant planets' orbits evolved through an episode of dynamical instability.  Several qualities of the solar system have recently been interpreted as evidence of this event transpiring within the first $\sim$100 Myr after the Sun's birth; around the same time as the final assembly of the inner planets.  In a series of recent papers we argued that such an early instability could resolve several problems revealed in classic numerical studies of terrestrial planet formation; namely the small masses of Mars and the asteroid belt.  In this paper, we revisit the early instability scenario with a large suite of simulations specifically designed to understand the degree to which Earth and Mars' formation are sensitive to the specific evolution of Jupiter and Saturn's orbits.  By deriving our initial terrestrial disks directly from recent high-resolution simulations of planetesimal accretion, our results largely confirm our previous findings regarding the instability's efficiency of truncating the terrestrial disk outside of the Earth-forming region in simulations that best replicate the outer solar system.  Moreover, our work validates the primordial 2:1 Jupiter-Saturn resonance within the early instability framework as a viable evolutionary path for the solar system.  While our simulations elucidate the fragility of the terrestrial system during the epoch of giant planet migration, many realizations yield outstanding solar system analogs when scrutinized against a number of observational constraints.  Finally, we highlight the inability of models to form adequate Mercury-analogs and the low eccentricities of Earth and Venus as the most significant outstanding problems for future numerical studies to resolve.

\end{abstract}

\section{Introduction}

Giant gaseous planets form precipitously within proto-planetary disks that dissipate on timescales of just a few Myr \citep{mizuno80,haisch01,mamajek08,pasucci09}.  In the standard paradigm, conditions in the solar nebula were appropriate to produce a sequential system of appropriately massed giant planet cores directly via pebble accretion \citep[e.g.:][]{morb_nesvory_12,lambrechts12,chambers14,levison15_gp} or gravitational collapse \citep[e.g.:][]{boss97,mayer02}.  In addition to feeding the outer planets' massive envelopes, the nebular gas conspired to force the giant planets into a tight chain of mutually resonant orbits via powerful gravitational torques \citep[e.g.:][]{masset01,morbidelli07,pierens08,zhang10,dangelo12}.  The Nice Model \citep{Tsi05,gomes05,mor05,nesvorny12} argues that a number of observed dynamical structures in the solar system \citep[e.g.:][]{levison08,morby09,nesvorny13,nesvorny15a,nesvorny15b,roig15} are best explained by the dissolution of this resonant chain through an epoch of dynamical instability.  Among other qualities, the existence of irregular satellites around all four giant planets \citep{nesvorny14a,nesvorny14b} strongly favors an epoch of planetary encounters having occurred sometime in the solar system's past.  

While a full review of the Nice Model is beyond the scope of this manuscript \citep[see, for example:][]{nesvorny18_rev,clement18,clement21_instb}, it is important to understand that the current consensus version invokes the existence of either one or two additional ice giants \citep[][note that we only consider cases with five primordial giant planets in this work]{nesvorny11}.  The ejection of these planets in successful numerical simulations of the highly-stochastic instability serves to both maximize the probability of retaining four outer planets \citep{nesvorny12}, and minimize the time powerful resonances with Jupiter and Saturn inhabit certain regions of the inner solar system \citep[e.g.:][]{bras09,minton10,walshmorb11,roig15,roig16}. 

Early investigations of the Nice Model advocated for a very specific timing of the event \citep{gomes05,levison11,deienno17}: $\sim$3.9 billion years ago in order to provide a natural trigger for the late heavy bombardment \citep[a spike in cratering rates in the inner solar system inferred via basin ages determined from the Apollo samples:][]{tera74}.  However, modern isotopic dating techniques \citep[e.g.:][]{norman06,grange13,merle14,mercer15,boehnke16} and new high-resolution imagery of the Moon's surface seem to imply a smooth decline in cratering rather than a terminal onslaught of impacts \citep{zellner17}.  In response to these revelations, dynamical investigations have increasingly sought to understand how varying the instability's precise timing affects the ability of models to match important observational and geochemical constraints.  Several recent studies in this mold have convincingly argued that particular aspects of the solar systems favor an early ($t\lesssim$ 100 Myr) instability.  These include the survival of Jupiter's Patroclus-Menoetius binary trojan pair \citep{nesvorny18}, the dichotomous inventories of highly siderophile elements incorporated in the mantles of the Earth and Moon following the formation of their cores \citep{morb18,brasser20}, resetting ages of various inner solar system meteorites \citep{mojzsis19}, collisional families in the asteroid belt with inferred ages $\gtrsim$4.0 Gyr \citep{delbo17,delbo19}, and the fact that late instabilities are highly improbable from a dynamical standpoint \citep{quarles19,ribeiro20}.  Thus, a diffuse consensus has developed over the past several years in support of the instability's occurrence within the first $\sim$100 Myr after nebular gas dispersal.

An early instability implies that the event occurred around the same time as the Moon-forming impact \citep[$t\simeq$ 30-100 Myr, e.g.:][]{earth}, and by extension coincident with the late stages of terrestrial planet formation \citep{wood05,kleine09,rudge10,kleine17}.  This is potentially advantageous as the delicate orbits of the fully formed terrestrial planets (namely those of Mercury and Mars) are easily destabilized by the Nice Model instability \citep{bras09,agnorlin12,bras13,kaibcham16}.  In the classical model of terrestrial planet formation \citep{wetherill80,wetherill91,chambers98} the inner planets collisionally accrete from an ocean of $\sim$0.01-0.1 $M_{\oplus}$ embryos \citep{koko_ida_96,koko_ida_98,koko_ida00,chambers06} and smaller, $D\simeq$ 10-1000 km planetesimals \citep{johansen07,morby09_ast,delbo17}.  If the terrestrial forming disk extends to the asteroid belt's outer edge \citep[consistent with the minimum mass solar nebula:][]{mmsn,hayashi81}, the resulting model-generated Mars analogs consistently possess masses similar to those of Earth and Venus \citep[an order of magnitude more massive than the real planet:][]{chambers01,ray09a}.  Similarly, massive planets in the asteroid belt are common outcomes in classic studies of terrestrial planet formation \citep{chamb_weth01}.  A straight-forward resolution to these problems involves restricting the amount of material available for Mars' accretion by either truncating the disk's outer edge \citep{wetherill78,agnor99,morishima08,hansen09} or altering its structure \citep{chambers_cassen02,iz14,izidoro15}.  These initial conditions might be explained by either radial variances in the efficiency of planetesimal formation \citep{levison15,draz16,ray17sci}, or the two-phased inward-outward migration of Jupiter and Saturn during the gas-disk phase \citep[e.g.: the Grand Tack model:][]{walsh11,pierens11,jacobson14,brasser16}.  For recent reviews of these models see: \citet{morb12}, \citet{izidoro18_book_review} and \citet{ray18_rev}.

An alternative solution to the small Mars problem relies on the influence of Jupiter and Saturn's eccentric forcing \citep{ray09a,lykawaka13,bromley17}.  \citet{clement18} studied the effects of an unusually early Nice Model ($t\lesssim$ 10 Myr) on the forming terrestrial planets and found reasonable solar system analogs result regularly in models where the instability excites the eccentricities of Jupiter and Saturn before Mars' mass exceeds its modern value.  By stunting Mars' growth in this manner, the instability essentially sets the planet's geological growth timescale; thus providing a natural explanation for the Mars' rapid inferred accretion time \citep{mars,Dauphas11,kruijer17_mars}.  Subsequent investigations by \citet{deienno18} and \citet{clement18_ab} leveraging high-resolution simulations of the asteroid belt found the early instability scenario to be broadly consistent with the belt's low-mass and dynamically excited state.  In particular, adequately exciting and depleting a primordially massive belt \citep[e.g.:][]{petit01} necessitates an instability strong enough to significantly perturb or destroy the system of fully formed terrestrial planets \citep{agnorlin12,kaibcham16}; thus axiomatically requiring an early Nice Model.  However, as a consequence of this violent dynamical event the final Earth and Venus analogs formed in early instability simulations \citep{clement18,nesvorny21_tp} typically possess eccentricities and inclinations that are too large.  \citet{clement18_frag} found marginally improved outcomes by considering the tendency of collisional fragments and debris to damp the forming planets' orbits \citep{chambers13,walsh16}, however the efficiency of this process is highly dependent on the numerical implementation \citep[for an opposing view see:][]{deienno19}.  Nevertheless, reconciling the dynamically cold orbits and, to a lesser degree, the compact radial separation of Earth and Venus remains a major shortcoming of terrestrial planet formation models in general \citep{ray18_rev}.

In this manuscript we return to the early instability framework of \citet{clement18} and \citet{clement18_frag}, henceforth Paper I and Paper II, respectively, with new simulations incorporating a number of important modifications.  In particular, several new developments in our understanding of the solar system's early evolution implore us to reexamine the early instability scenario:

\begin{enumerate}
	\item The instability itself is inherently stochastic, and only a few percent of numerical realizations yield giant planet configurations broadly akin to the real outer solar system \citep{bat10,nesvorny12,batbro12,deienno17}.  Thus, only a small number of the original simulations in Paper I and Paper II produced reasonable outer solar system analogs.  Indeed, many of the less successful terrestrial results were derived from simulations that under-excited the eccentricities of Jupiter and Saturn \citep[a facet of the outer solar system that is challenging to replicate with N-body simulations:][see below]{nesvorny12,clement21_instb}.  In this paper, we develop a new pipeline that substantially increases our sample of appropriate giant planet evolutions with the aim of more concretely understanding the dependence of the early instability scenario's viability on the eccentricity excitation of the gas giants.  \citet{nesvorny21_tp} considered a similar question by modeling successful instabilities via cubic interpolation of simulation outputs and concluded that the ability of the scenario to replicate Mars' mass is not related to the particular evolution of Jupiter and Saturn.  However, the authors did not control for the instability's timing, thus making it difficult to disentangle the underlying cause of variations in statistical outcomes between different batches of simulations (discussed further in \ref{sect:e55}).
	\item The investigations in Paper I, Paper II and \citet{nesvorny21_tp} exclusively considered Nice Model scenarios where Jupiter and Saturn originate in a 3:2 MMR (mean motion resonance) on circular orbits \citep[e.g.:][]{morbidelli07,pierens08}.  While such initial conditions have demonstrated consistent success when scrutinized against certain small body constraints \citep[see][for a relevant review]{nesvorny18_rev}, it is systematically challenging to adequately excite Jupiter's eccentricity ($e_{J}$) without over-exciting Saturn's eccentricity ($e_{S}$) and driving its semi-major axis ($a_{S}$) into the distant solar system \citep{clement21_instb}.  This shortcoming is unfortunate given the apparent correlation between the gas giants' eccentricities and Mars' mass \citep{ray09a,clement18}.  \citet{pierens14} found that Jupiter and Saturn's capture in the 2:1 MMR is possible given a specific combinations of assumed disk parameters.  Of particular relevance to the topic of terrestrial planet formation and the small Mars problem, the gas giants carve out larger gaps in the nebular gas when locked in the 2:1 resonance; thus allowing them to attain inflated eccentricities prior to the instability.  \citet{clement21_instb} studied the outcomes of these 2:1, high-eccentricity instabilities statistically and noted markedly improved success rates compared to the primordial 3:2 MMR in terms of matching Jupiter's eccentricity without driving Saturn's semi-major axis into the distant solar system.  As the giant planets' resonant perturbations in the terrestrial region occur over a more restricted radial range in these new evolutions, it is important to thoroughly study the implications of the primordial 2:1 Jupiter-Saturn resonance on the early instability scenario.
	\item The initial conditions for the giant impact phase supposed throughout the literature (as well as in Papers I and II) are loosely based on the outcomes of semi-analytic investigations of runaway growth \citep[e.g.:][]{koko_ida_96,koko_ida_98,koko_ida00,chambers06}.  However, the computational challenge of directly resolving proto-planet growth from $\sim$100 km planetesimals makes it difficult to infer the precise embryo and planetesimal distributions in the terrestrial disk around the time of nebular gas dissipation.  Consequently, terrestrial planet formation studies tend to either distribute equal-mass embryos throughout the terrestrial disk \citep[e.g.:][]{chambers01,obrien06}, or assign embryos masses that are proportional to the analytic isolation mass \citep[e.g.:][]{raymond04,ray06,ray09a}.  Advances in computing power have recently made high-resolution N-body models of runaway growth throughout the terrestrial disk feasible \citep[e.g.:][]{morishma10,carter15,walsh19,wallace19,clement20_psj,woo21}.  This presents a novel opportunity to test the viability of the early instability scenario with embryo and planetesimal distributions derived from scratch.  Here, we utilize outputs from \citet{walsh19} and \citet{clement20_psj} around the time of nebular dispersal when constructing our terrestrial disks. 
	\end{enumerate}

It is worth mentioning that contemporary terrestrial planet formation models are incapable of consistently generating Mercury analogs of the appropriate mass \citep{chambers01,obrien06,lykawka17,lykawka19}, composition \citep{hauck13,nittler17,jackson18}, and radial offset from Venus \citep{clement19_merc}.  While we do not neglect Mercury in the analysis sections of this paper (we present a particularly interesting Mercury analog in section \ref{sect:merc}), we also do not explicitly modify our simulations to boost the likelihood of forming Mercury \citep{lykawka17,lykawka19,clement21_merc3,clement21_merc2}.  Thus, the primary goals of this paper are to validate the viability of the primordial 2:1 Jupiter-Saturn resonance \citep{pierens14,clement21_instb} and the disk-evolved embryo and planetesimal distributions from \citet{walsh19} and \citet{clement20_psj} within the early instability framework of Paper I.

\section{Methods}

\subsection{Numerical Simulations}

Our numerical approach largely follows the methodology established in Paper I.  To ensure the instability triggers at a predetermined time (i.e.: in conjunction with a specific evolutionary state of the terrestrial disk) we integrate our terrestrial disks (section \ref{sect:methods_tp}) and giant planet configurations (section \ref{sect:methods_inst}) separately before combining both sets of bodies into a single simulation.  We construct these terrestrial disk inputs directly using time-outputs of interest from high-resolution simulations of planetesimal accretion and runaway growth in \citet{walsh19} and \citet{clement20_psj}.  To minimize interpolations, embryo populations are derived exactly from all objects with $M>$ 0.01 $M_{\oplus}$ (around the mass of the Moon).  Planetesimal distributions are inferred by sampling from the remaining particles' orbital distributions.

We generate systems of resonant giant planets \citep[][described further in section \ref{sect:methods_inst}]{nesvorny12,clement21_instb} with fictitious forces designed to mimic gas disk interaction \citep[e.g.:][]{lee02}.  We then integrate the resultant resonant chains in the presence of an external disk of primordial Kuiper Belt Objects \citep[KBOs, e.g.:][]{fer84,hahn99,levison08,nesvorny15a,nesvorny15b,quarles19} until two giant planets experience a close encounter within three mutual Hill radii.  In the vast majority of cases this initial close approach is indicative of an imminent instability (the majority of simulations tested in this manner in Paper I experienced an instability within 100 Kyr).  At this point, we combine the giant planets, surviving KBOs, terrestrial embryos and planetesimals (section \ref{sect:methods_tp}) in one single simulation.  As in Paper I and Paper II, all of our terrestrial planet formation computations last for 200 Myr, utilize a 6 day time-step, and leverage the \textit{Mercury6} Hybrid integrator \citep{chambers99}. The giant planets and terrestrial embryos are treated as fully active particles (i.e.: they both perturb, and experience the gravity of all the other objects in the simulation).  Conversely, KBOs and planetesimals only feel the gravitational effects of the planets and embryos.  Objects are considered ejected from the system at heliocentric distances of 100 au, and removed via merger with the Sun if they attain perihelia less than 0.1 au \citep[e.g.:][]{chambers01}.

\subsection{Simulation pipeline}
\label{sect:meth:restart}

Our giant planet instability models (section \ref{sect:methods_inst}) yield a broad range of final system outcomes \citep{nesvorny12,clement21_instb} in terms of the surviving number of giant planets ($N_{GP}$), their semi-major axes, eccentricities and inclinations.  While the terrestrial planets' orbital evolution is not particularly sensitive to the peculiar dynamics of Uranus and Neptune, the ultimate semi-major axes and eccentricities of the gas giants are important for determining the fate of the inner planets \citep{levison03,morby09,bras09,ray09a}.  In addition to establishing the precise locations of dominant MMRs in the inner solar system, the Jupiter-Saturn orbital spacing (commonly parameterized by the planets' orbital period ratio: $P_{S}/P_{J}$) sets the dominant eccentric precession eigenfrequencies $g_{5}$ and $g_{6}$ \citep{nobili89,laskar90} that are responsible for the powerful $\nu_{5}$ and $\nu_{6}$ secular resonances stretching throughout the inner solar system \citep{morby91,morby91b,michel97}.  As the instability evolution of the $\nu_{6}$ resonance largely sculpts the primordial asteroid belt into its modern form \citep[e.g.:][]{walshmorb11,minton11,deienno18,clement20_mnras}, and the $\nu_{5}$ resonance is chiefly responsible for chaotic evolutionary trajectories of the planet Mercury \citep{laskar09,batygin15b} in the modern solar system, it is extremely important that our simulations consistently replicate $P_{S}/P_{J}$.  Additionally, depletion in the Mars and asteroid belt region is particularly sensitive to Jupiter's final eccentricity \citep[commonly quantified by the amplitude of the $g_{5}$ eccentric mode in its orbit: $e_{55}=$ 0.044:][]{ray09a,morby09,nesvorny12,clement18}.  Thus, replicating Jupiter's modern eccentricity in this manner represents another key constraint for our models.

In Paper I we ran all simulations to completion (regardless of the post-instability values of $P_{S}/P_{J}$ and $e_{55}$), and analyzed simulations with adequate final giant planet orbits separately.  To save compute time in Paper II, we terminated simulations when $P_{S}/P_{J}$ exceeded 2.8 \citep[i.e.: the runs were considered failures:][]{nesvorny12}.  While this increased the sample of solar system-like giant planet configurations to first order, the majority of these instabilities still finished with under-excited $e_{55}$ values.  To bolster the sample of systems finishing with $e_{55}\simeq$ 0.044 and $P_{S}/P_{J}<$ 2.5 in our current investigation we monitor both the Jupiter-Saturn period ratio and Jupiter's time-averaged eccentricity ($e_{J}$: a reasonable proxy for $e_{55}$) throughout the integration with a rolling 100 Kyr time window.  Simulations are restarted with a new unique set of initial conditions (see sections \ref{sect:methods_inst} and \ref{sect:methods_tp}) if any of the following criteria are met:

\begin{enumerate}
	\item More than 1 Myr elapsed without an instability as determined by either an ice giant ejection or a step-change in $P_{S}/P_{J}$.
	\item $P_{S}/P_{J}>$ 2.5 within the first 200 Kyr after the instability's onset.
	\item $e_{J}<$ 0.03 within the first 200 Kyr after the instability excites $e_{J}$.
	\item $P_{S}/P_{J}>$ 2.8 any time after the instability.
	\item $e_{J}<$ 0.01 any time after the instability (approximately half of Jupiter's modern minimum eccentricity).
\end{enumerate}

In practicality this is accomplished by dedicating a prescribed number of compute cores to a given simulation batch for around 8 months.  Each core continuously restarts new simulations until a ``good'' instability occurs, which is then saved for analysis after reaching $t=$ 200 Myr.  Naturally this methodology still generates a range of evolutions, some of which are not the best solar system analogs.  However, it substantially increases the number of solar system-like final giant planet configurations while still producing a range of $P_{S}/P_{J}$ and $e_{55}$ outcomes that are useful for understanding trends.  As we seek to maximize the number of successful instabilities while minimizing compute time in this manner, we do not incorporate a collisional fragmentation algorithm \citep[e.g.:][]{leinhardt12,stewart12,genda12,chambers13} as in Paper II.  However, it should be noted that this choice likely inhibits the ability of Earth and Venus to form on dynamically cold orbits (discussed further in section \ref{sect:amd}).

\subsection{Instability Models}
\label{sect:methods_inst}

\begin{table*}
\centering
\begin{tabular}{c c c c c c c c c c}
\hline
Name  & $N_{Pln}$ & $e_{J,o}$ & $e_{S}$ & $M_{disk}$ & $\delta$r & $r_{out}$ & $a_{nep}$ & Resonance Chain & $M_{ice}$\\
& & & & ($M_{\oplus}$) & (au) & (au) & (au) & & ($M_{\oplus}$)\\
\hline
3:2 & 5 & 0.0 & 0.0 & 35 & 1.5 & 30 & 17.4 & 3:2,3:2,3:2,3:2 & 16,16,16 \\
2:1 & 5 & 0.05 & 0.05 & 20 & 1.5 & 30 & 18.6 & 2:1,4:3,3:2,3:2 & 8,16,16 \\
\hline
\end{tabular}
\caption{Table of giant planet initial resonant configurations.  The columns are: (1) the name of the simulation set, (2) the number of giant planets, (3-4) the initial eccentricities of Jupiter and Saturn, (5) the mass of the planetesimal disk exterior to the giant planets, (6) the distance between the outermost ice giant and the planetesimal disk’s inner edge, (7) the location of the disk's outer edge, (8) the semi-major axis of the outermost ice giant, (9) the resonant configuration of the giant planets starting with the Jupiter-Saturn resonance, and (10) the masses of the ice giants from inside to outside.}
\label{table:gp}
\end{table*}

\begin{figure*}
	\centering
	\includegraphics[width=.495\textwidth]{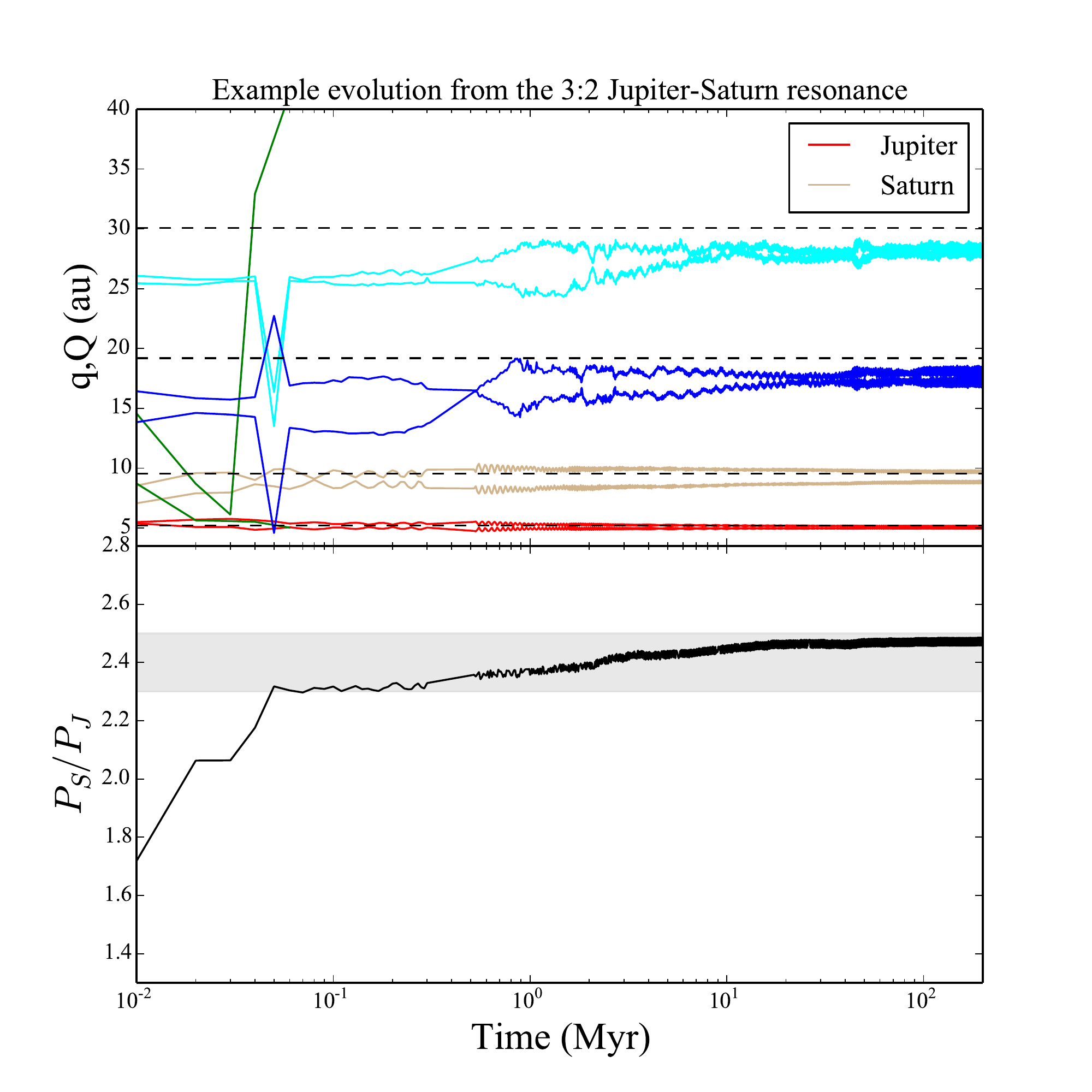}
	\includegraphics[width=.495\textwidth]{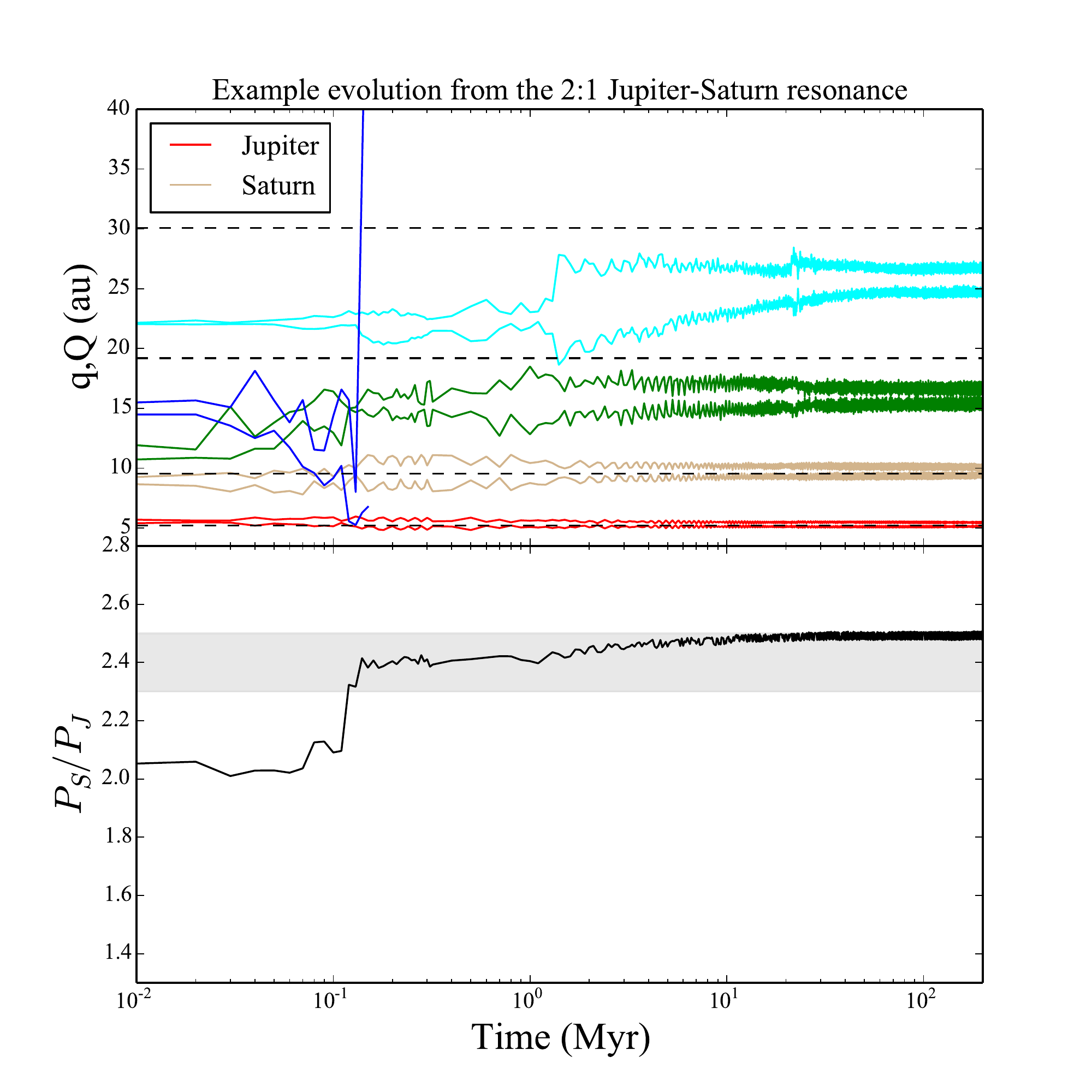}
	\caption{Example instability evolutions from our contemporary simulation suite beginning with Jupiter and Saturn in a 3:2 MMR \citep[left panel, e.g.:][]{nesvorny12} versus the 2:1 MMR \citep[right panel, e.g.:][]{clement21_instb}.  The final e55 values for both simulations are $\sim$0.030.  The top panel plots the perihelion and aphelion of each planet over the length of the simulation.  The bottom panel shows the Jupiter-Saturn period ratio.  The horizontal dashed lines in the upper panel indicate the locations of the giant planets' modern semi-major axes.  The shaded region in the middle panel delimits the range of 2.3 $<P_{S}/P_{J}<$ 2.5.}
	\label{fig:gp_ex}
\end{figure*}

Our work tests two separate instability models: one where Jupiter and Saturn originate in a 3:2 MMR \citep[a 3:2,3:2,3:2,3:2\footnote{Note that resonant chains are reported as the ratio of successive planets' period ratios with increasing semi-major axis} resonant chain demonstrated successful in][]{nesvorny12} and one where the planets begin locked in a primordial 2:1 MMR with inflated eccentricities \citep[a 2:1,4:3,3:2,3:2 chain favored in the analysis of][]{clement21_instb}.  The initial conditions for these instability models are summarized in table \ref{table:gp}, and example evolutions from our contemporary simulation suite are plotted in figure \ref{fig:gp_ex}.  The major motivation for testing both the 3:2 and 2:1 Jupiter-Saturn configurations in this paper rather than focusing on the 3:2 as in Paper I and Paper II is to understand the consequences of primordial eccentricity excitation within the 2:1 \citep[a result of the planets' carving larger gaps within the nebular gas:][]{pierens14}.  The complete methodology for generating eccentric resonant chains is described in \citet{clement21_instb}.  In short, we migrate the planets into the desired resonant chain by incorporating forced migration ($\dot{a}$) and eccentricity damping ($\dot{e}$) terms in the equations of motion.  Once in resonance, we reduce the magnitude of the eccentric damping (in some cases reversing its sign) on Jupiter and Saturn to artificially pump their eccentricities.

After the chains are assembled, we surround the resonant giant planets with a disk of 1,000 equal-mass KBOs conforming to the parameters provided in table \ref{table:gp}.  In all cases, the disks' radial surface density profile is proportional to $r^{-1}$, the inner disk edge is offset from the outermost planet by 1.5 au \citep{nesvorny12,quarles19}, and the outer edge is set at 30.0 au \citep{gomes03}.  Eccentricities and inclinations for the planetesimals are drawn from near-circular distributions as in Paper I, and the remaining orbital elements are selected randomly from uniform distributions of angles.  We then integrate a large number of instability simulations (without terrestrial disks) with the $Mercury6$ Hybrid integrator and a 50 day time-step.  KBOs do not feel the gravitational perturbations from one another in our simulations.  Simulations are stopped when the code flags a close encounter between two planets within three mutual Hill Radii.  Through this process we generate a large sample of unique outer solar system architectures on the verge of instability to select from when generating unique initial conditions for our terrestrial planet formation simulations described below.

\subsection{Terrestrial Disks}
\label{sect:methods_tp}

We utilize outputs from \citet{walsh19} and \citet{clement20_psj} as inputs for our terrestrial planet formation simulations.  Each study models collisional evolution and runaway growth in the inner solar system throughout the gas disk phase beginning from $D\sim$ 10-100 km planetesimals.  The following subsections provide brief synopses of each author's methodology.

\begin{table*}
\centering
\begin{tabular}{c c c c c c c c c}
\hline
Set & $a_{in}$-$a_{out}$ (au) & $N_{emb}$ & $N_{pln}$ & $M_{emb,tot}$ ($M_{\oplus}$) & $M_{pln,tot}$ ($M_{\oplus}$)  & $t_{instb}$ (Myr) & $N_{sim}$\\
\hline
C18/control & 0.5-4.0 & 100 & 1000 & 2.5 & 2.5 & N/A & 50 \\
C18/3:2/1Myr & 0.5-4.0 & 100 & 1000 & 2.5 & 2.5 & 1 & 15 \\
C19/control & 0.5-4.0 & 100 & 1000 & 2.5 & 2.5 & N/A & 100 \\
C19/3:2/1Myr & 0.5-4.0 & 100 & 1000 & 2.5 & 2.5 & 1 & 17 \\
C20/control & 0.48-4.0 & 25 & 500 & 2.18 & 2.43 & N/A & 25 \\
\hline
C20/3:2/0Myr & 0.48-4.0 & 25 & 1000 & 2.18 & 2.43 & 0 & 25 \\
C20/2:1/0Myr & 0.48-4.0 & 25 & 1000 & 2.18 & 2.43 & 0 & 35  \\
C20/3:2/5Myr & 0.48-4.0 & 23 & 954 & 2.25 & 2.33 & 5 & 20 \\
C20/2:1/5Myr & 0.48-4.0 & 23 & 954 & 2.25 & 2.33 & 5 & 43 \\
WL19/3:2/0Myr & 0.7-3.0 & 20 & 1000 & 0.45 & 2.45 & 0 & 27  \\
WL19/2:1/0Myr & 0.7-3.0 & 20 & 1000 & 0.45 & 2.45 & 0 & 27 \\
WL19/3:2/5Myr & 0.7-3.0 & 23 & 700 & 1.01 & 1.71 & 5 & 21 \\
WL19/2:1/5Myr & 0.7-3.0 & 23 & 700 & 1.01 & 1.71 & 5 & 50  \\
WL19/3:2/15Myr & 0.7-3.0 & 16 & 401 & 1.18 & 0.98 & 15 & 43 \\
WL19/2:1/15Myr & 0.7-3.0 & 16 & 401 & 1.18 & 0.98 & 15 & 65 \\
\hline	
\end{tabular}
\caption{Summary of initial conditions for complete sets of terrestrial planet formation simulations.  The columns are as follows: (1) the name of the simulation set (note that the acronyms C20 and WL19 denote the terrestrial disks inferred from \cite{clement20_psj} and \cite{walsh19}, respectively, while the designators C18 and C19 represent simulations taken from Paper I and Paper II for comparison), (2) the inner and outer edges of the terrestrial forming disk, (3-4) the total number of embryos and planetesimals, (5-6) the total mass of the embryo and planetesimal components, (7) the instability timing in Myr, and (8) the total number of integrations comprising the set.}
\label{table:ics2}
\end{table*}

\subsubsection{\citet{walsh19}}
\label{sect:wl19}

We extract outputs from the nominal case presented in \citet{walsh19} to derive our terrestrial disks denoted as \textit{WL19} in table \ref{table:ics2}.  The authors utilize the \textit{LIPAD} numerical integration code \citep{lipad} that facilitates the modeling of large numbers of particles with various sizes by treating small objects as ``tracer'' particles.  While more massive objects in the simulation are integrated with a direct N-body approach, the tracer particles' collisional interactions are handled statistically.  The nominal \textit{LIPAD} simulation used in our work considers a terrestrial disk initially composed of $D\simeq$ 60 km planetesimals.  The disk extends from 0.7-3.0 au and contains 3.32 $M_{\oplus}$ of solid material distributed with an initial surface density profile that falls off radially as $r^{-3/2}$.  Jupiter and Saturn are included in the simulation as 1 $M_{\oplus}$ cores at 3.5 and 6.0 au up until $t=$ 4 Myr when they are moved to 5.0 and 9.5 au and inflated to their modern masses \citep[roughly analogous to our 2:1 pre-instability configuration:][]{clement21_instb}. The authors also incorporate a gas disk model \citep[e.g.:][]{tanaka04} based on the nominal minimum mass solar nebula.  Gas decays uniformly in space and exponentially in time with $\tau=$ 2 Myr.  

The first simulation output we utilize is the $t=$ 2 Myr output, which we refer to as either time zero for the remainder of this text.  This allows us to make comparisons with the terrestrial disk models inferred from \citet[][subsequent section]{clement20_psj} that utilize a value of $\tau=$ 3 Myr by standardizing time zero as the first point when data is extracted from the runaway growth model (loosely related to the nebular gas dispersal time).  Thus, the WL19 models in table \ref{table:ics2} investigating ``instability times'' of 0, 5 and 15 Myr correlate with the $t=$ 2, 7 and 17 Myr time outputs from \citet{walsh19}, respectively.

\begin{figure}
	\centering
	\includegraphics[width=.5\textwidth]{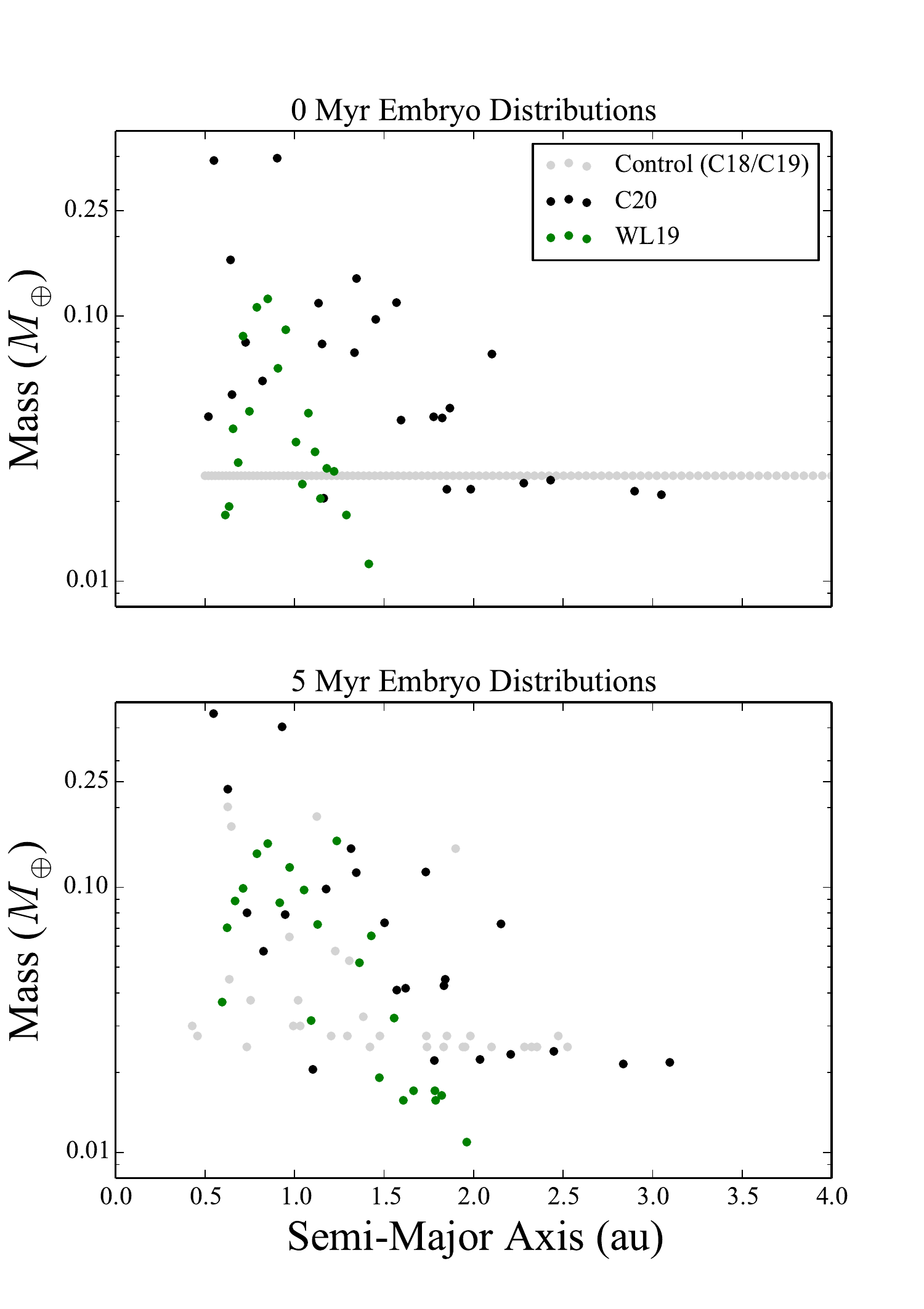}
	\caption{Comparison of embryo distributions for various disks tested in this manuscript.  The acronyms C20 (black points) and WL19 (green points) denote the terrestrial disks inferred from \cite{clement20_psj} and \cite{walsh19}, respectively, while the designators C18 and C19 (grey points) represent simulation results taken from Paper I and Paper II for comparison.  The top panel represents the state of each disk at $t=$ 0 Myr; the gas dispersal time.  Conversely, the bottom panel plots the conditions of the same disks at $t=$ 5 Myr.  The C18/C19 points in the top panel are example initial conditions for control simulations from the respective works, where as the similar points in the bottom panel are an example embryo distribution from a C19 control simulation at $t=$ 5 Myr (for an example of the state of the terrestrial disk prior to an instability triggered at the 5 Myr point in that work).}
\label{fig:kev_compare}
\end{figure}

\subsubsection{\citet{clement20_psj}}
\label{sect:c20}

Our simulations incorporating terrestrial disks denoted as \textit{C20} (table \ref{table:crit}) use outputs from GPU-accelerated simulations reported in \citet{clement20_psj}.  These computations leverage the \textit{GENGA} integrator \citep{genga} that is designed to speed-up simulations by performing certain sets of calculations in parallel.  The effects of nebular gas interactions are included using the numerical implementation of \citet{morishma10}.  The initial disk is modeled utilizing a multi-annulus approach for the first 1 Myr of the total simulation time.  During this evolutionary phase, the disk is split into different overlapping annular sections that are integrated independently until the total particle number drops low enough for annuli to be merged.  The initial disk spans the range of 0.48-4.0 au and possesses a total mass, $M_{tot}=$ 5.0 $M_{\oplus}$ of terrestrial forming material with a radially dependent surface density profile proportional to $r^{-3/2}$.  All particles have the same initial size of $D\simeq$ 200 km.  At $t=$ 1 Myr all annuli are combined into a single simulation containing 43,608 particles in the terrestrial-forming disk.  At this stage, 8.0 $M_{\oplus}$ versions of Jupiter and Saturn are added in a 3:2 resonance \citep[consistent with the initial conditions of our 3:2 instability configurations:][]{nesvorny12,deienno17}. At $t=\tau=$ 3 Myr, Jupiter and Saturn begin to grow logarithmically such that they reach 95$\%$ of their modern masses at $t=$ 6 Myr. 

We study the 3 and 8 Myr outputs from these simulations.  For the remainder of the text we refer to these as instability times of 0 and 5 Myr, respectively (the timing of the instability related to $\tau$).  Unlike for the WL19 \textit{LIPAD} simulations, the embryos in the Mars region in the C20 GPU computations grow beyond the modern mass of Mars after around $t=$ 5 Myr \citep[the reasons for these differences are analyzed in detail in][]{clement20_psj}.  Thus, we do not investigate an instability time of $t=$ 15 Myr with the C20 disks as Mars has already grown too large for its mass to be effectively limited by the instability \citep{clement18,nesvorny21_tp}.  Later instability times might be viable if the largest embryo in the Mars-region is lost (e.g.: via collision with Earth or Venus), thus stranding a smaller embryo as the Mars analog.  However, to avoid overly-expanding our tested parameter space, we reserve the exploration of this possibility for future work.
\subsubsection{Disk Interpolation}

We import our embryo distributions (table \ref{table:ics2}) directly from all terrestrial objects with $M\geq$ 0.01 $M_{\oplus}$ in the relevant simulation output files.  Figure \ref{fig:kev_compare} plots these distributions for the 0 and 5 Myr instability cases, along with the ``classic'' disk conditions \citep{chambers98,ray09a} supposed in control simulations from Paper I and Paper II used as comparison cases throughout the analysis sections of this paper (see section \ref{sect:control}).  The various instability times and different disk models interrogated in this work should be interpreted as different possible evolutionary states the terrestrial disk might have attained around the time of the instability's onset (in terms of the total number of embryos and their cumulative masses).  In this paradigm, it is important to note that the major difference between our WL19 and C20 disks is the more advanced evolutionary nature of the C20 models.  This is partially a consequence of the fact that we extract data from the original C20 \textit{GENGA} simulations at a later epoch ($t=$ 3 Myr) than for the WL19 \textit{LIPAD} models ($t=$ 2 Myr).  Moreover, the \citet{walsh19} simulations incorporate algorithms designed to account for the effects of collisional fragmentation, thus potentially limiting the efficiency of runaway growth \citep{chambers13}.  Indeed, embryos in the Earth and Venus-forming regions of our C20 models already possess masses of $\sim$0.3-0.4 $M_{\oplus}$ around the time of gas disk dispersal (figure \ref{fig:kev_compare}).  In all simulations, embryos interact gravitationally with all other objects in the system.

We generate planetesimal populations for our simulations by randomly sampling the mass-weighted distributions of objects with $M<$ 0.01 $M_{\oplus}$ in the WL19 and C20 output files such that the in-situ mass distribution (largely consistent with the original $r^{-3/2}$ profile in both cases) is roughly maintained \citep{birnstiel12}.  Each time a simulation is re-started (i.e.: after an unsuccessful instability as described in section \ref{sect:meth:restart}) a new planetesimal population is selected and paired with a new, randomly selected giant planet configuration.  Thus, each simulation combines a common embryo population (depending on the specific simulation set: figure \ref{fig:kev_compare}) with a unique outer solar system and a unique planetesimal population.  In all simulations, planetesimals only feel the gravitational effects of the embryos and giant planets.  It should be noted here that our method of extracting simulation outputs from the WL19 \textit{LIPAD} and C20 \textit{GENGA} simulations to pair with separate outer solar system configurations imposes an artificial step change on the giant planets' semi-major axes and masses.  While this is a necessary simplification in our case, it limits the robustness of our results, and future endeavors should strive to more accurately model the transition from the nebular gas phase of evolution to the epoch of giant impacts.

We distribute 1,000 equal-mass planetesimals in our $t=$ 0 Myr instability simulations such that the total mass in planetesimals is equal to the total mass of all objects with $M<$ 0.01 $M_{\oplus}$ in the original \textit{GENGA} and \textit{LIPAD} simulations (column 6 of table \ref{table:ics2}).  Therefore, planetesimals in our integrations investigating WL19 disks have initial masses of 0.00245 $M_{\oplus}$, and those in C20 models possess masses of 0.00243 $M_{\oplus}$ ($M_{pln}$).  In order to make accurate comparisons between our simulations testing different instability times that are not complicated by differences in individual planetesimal masses, we maintain the values of $M_{pln}$ from our 0 Myr instability runs in the 5 and 15 Myr instability sets.  Thus, disks constructed for the 5 and 15 Myr instability simulations necessarily possess fewer planetesimals as a result of their more advanced evolutionary states.  This also explains the broad differences between the total number of fully evolved systems we produce from each set of initial conditions (last column of table \ref{table:ics2}).  Simulations investigating 5 and 15 Myr instability delays require less compute time to complete than the 0 Myr batches by virtue of beginning with fewer particles.  Similarly, our 2:1 Jupiter-Saturn configurations yield successful instabilities more regularly \citep{clement21_instb}, and thus require fewer ``restarts'' to generate a fully evolved system.  It follows then that our WL19/2:1/15Myr set generates the largest sample of evolved systems.  An alternative means of interpolating planetesimal populations would be to assign each object a different mass that is commensurate with the original simulations' resultant size frequency distribution, and we plan to explore the effects of different planetesimal distributions further in future work.

\subsubsection{Reference Cases}
\label{sect:control}

We reference simulations of the classic model of terrestrial planet formation \citep[both with and without a giant planet instability:][]{chambers98,ray09a,clement18} throughout our manuscript.  These simulation batches are summarized at the top of table \ref{table:ics2}.  Disks denoted $C18$ (figure \ref{fig:kev_compare}) are derived from Paper I, and incorporate 100 equal-mass embryos and 1,000 planetesimals, a surface density profile proportional to $r^{-3/2}$, and a total disk mass of 5.0 $M_{\oplus}$.  Disks labeled $C19$ (Paper II) are identical to the C18 disks, however these simulations are run using an algorithm that accounts for fragmenting and hit-and-run collisions \citep{leinhardt12,stewart12,chambers13}.  Otherwise, the numerical methodology for both control sets is identical to that described in this paper.  In sets labeled ``control'' cases, Jupiter and Saturn are included in their pre-instability, 3:2 MMR configuration for the entire 200 Myr integration.  Additionally, we reference simulations from each work (C18/3:2/1Myr and C19/3:2/1Myr) that include a giant planet instability (the same 3:2 case tested here: table \ref{table:gp}) at $t=$ 1 Myr; the most successful instability time that is common to both works.  While we only present instability simulations from these papers that finish with $P_{S}/P_{J}<$ 2.8 (as described in section \ref{sect:meth:restart}), it is important to note that these models possess a broader range of final giant planet eccentricities than in our contemporary simulations.  Finally, simulations designated C20/control study the C20/0Myr disks (section \ref{sect:c20}) without a giant planet instability model (Jupiter and Saturn are instead modeled in a 3:2 resonance).

\subsection{Success Criteria}
\label{sect:scs}

\begin{table*}
\centering
\begin{tabular}{c c c c c}
\hline
Code & Criterion &  Actual Value & Accepted Value & Justification\\
\hline
A & $a_{Mars}$ & 1.52 au & 1.3-2.0 au & Sunward of AB \\
A & $M_{Mars}$ & 0.107 $M_{\oplus}$ & $\geq0.05$,$<.3 M_{\oplus}$ & \citep{ray09a} \\
A1 & $M_{Mars}$ & 0.107 $M_{\oplus}$ & $\geq 0.0$,$<.3 M_{\oplus}$ & \citep{ray09a} \\
A,A1 & $M_{Venus}$ & 0.815 $M_{\oplus}$ & $>$0.6 $M_{\oplus}$ & Within $\sim25\%$\\
A,A1 & $M_{Earth}$ & 1.0 $M_{\oplus}$ & $>$0.6 $M_{\oplus}$ & Match Venus\\
B & $\tau_{Mars}$ & 1-10 Myr & $<$10 Myr & \citep{kleine09} \\
C & $\tau_{\oplus}$ & 50-150 Myr & $>$50 Myr & \citep{Dauphas11} \\
D & $M_{AB}$ & $\sim$ 0.0004 $M_{\oplus}$ & No embryos & \citep{chambers01} \\
E & $\nu_{6}$ ratio & & Not used here \\
F & $WMF_{\oplus}$ & & Not used here\\
G & AMD & 0.0018 & $<$0.0036 & \citep{ray09a} \\
\hline
\end{tabular}
\caption{Summary of success criteria for the inner solar system from Paper I and Paper II (a reproduction of table 3 from each respective work).  The rows are: (1) the semi-major axis of Mars, (2-4) The masses of Mars, Venus and Earth, (5-6) the time for Mars and Earth to accrete $90\%$ of their mass, (7) the final mass of the asteroid belt, (8) the ratio of asteroids above to below the $\nu_{6}$ secular resonance between 2.05-2.8 au, (8) the water mass fraction of Earth, and (9) the angular momentum deficit (AMD) of the inner solar system.}
\label{table:crit}
\end{table*}

We adopt the same success criteria (and alpha-numeric designators) utilized in Paper I and Paper II for consistency.  These constraints are summarized in table \ref{table:crit}.  Criterion \textbf{A} analyzes the bulk radial mass distribution of the final system of terrestrial planets.  In our classification algorithm, all objects with $M>$ 0.05 $M_{\oplus}$ are considered planets \citep[note that we largely neglect Mercury's formation:][]{clement19_merc,clement21_merc3,clement21_merc2}, and all other remaining terrestrial objects are referred to as left-over embryos and planetesimals.  To satisfy criterion \textbf{A}, a system must finish with exactly two planets (Earth and Venus analogs) with $m>$ 0.6 $M_{\oplus}$ and $a<$ 1.3 au, at least one planet (Mars analog) with $m<$ 0.3 $M_{\oplus}$ and 1.3 $<a<$ 2.0 au (approximately between Mars' modern pericenter and the inner edge of the asteroid belt), and no planets with $m>$ 0.3 $M_{\oplus}$ and $a>$ 1.3 au.  However, to first order, an instability simulation might still be successful if Mars finishes with $M<$ 0.05 $M_{\oplus}$, $a>$ 2.0 au, or if the system possess no Mars analog at all (see further discussion in Paper I).  For these reasons, and to prevent over-constraining our simulations with criterion \textbf{A}, we report these three types of systems (no Mars, Mars too small, Mars in asteroid belt) as successful with criterion \textbf{A1}, provided the Earth and Venus analogs still satisfy the constraints described above.

Criteria \textbf{B} and \textbf{C} scrutinize the relative formation timescales of Earth and Mars analogs.  As in Papers I and II, Earth analogs are defined as the largest object in each simulation with 0.85 $<a<$ 1.3 au and $m>$ 0.6 $M_{\oplus}$, and Mars analogs comprise the largest planets in each system with 1.3 $<a<$ 2.0 au, regardless of classification in terms of \textbf{A}.  Isotopic dating of lunar samples indicate that the final major accretion event on Earth (the Moon-forming impact) occurred $\sim$30-100 Myr after nebular gas dissipation \citep{yin02,wood05,earth,kleine09,rudge10,zube19}, although Earth's formation timescale and the timing of the Moon-forming giant impact is still a topic for debate \citep[e.g.:][]{barboni17,thiemens19}.  Conversely, analyses of the Martian meteorites suggest that Mars' formation was complete within just a few Myr of gas dispersal \citep{mars,Dauphas11,kruijer17_mars,costa20_mars}.  In order to assess a system's ability to reproduce these bifurcated accretion histories, criterion \textbf{B} requires that our Mars analogs accrete 90$\%$ of their final mass by $t=$ 10 Myr (relative to the gas dispersal time).  Conversely, to be successful in terms of criterion \textbf{C} an Earth analog must take longer than 50 Myr to reach 90$\%$ of its ultimate mass.

A detailed analysis of the dynamical evolution of the asteroid belt is beyond the scope of this manuscript \citep[see:][for studies of the early instability scenario's consequences in the belt]{deienno18,clement18_ab,clement20_mnras}.  However, a first order assessment of the instability's ability to adequately deplete the belt can be made by scrutinizing the left-over particles in the asteroid belt.  As planetesimals do not grow through collisions or feel the gravitational effects of one another in our simulations, they can be viewed as tracers of the more numerous planetesimals that actually populated the primordial belt.  For this reason, criterion \textbf{D} simply requires that our simulations finish with no surviving embryos or planets in the asteroid belt \citep[2.0 $<a<$ 4.0 au][]{chambers01,chamb_weth01} as these objects would fossilize gaps in the belt's orbital distribution that are not observed today \citep{petit01,obrien07,ray09a,izidoro16}.

In Paper I and Paper II we analyzed the orbital architectures of our remnant asteroid belts by requiring that systems possess more asteroids in the inner belt with inclinations below the $\nu_{6}$ secular resonance than above it \citep[criterion \textbf{E}, the ratio in the actual belt is $\sim$0.08:][]{walshmorb11}.  Recent work in \citet{clement20_mnras} demonstrated that the giant planets' residual migration phase (i.e.: migration after the instability) largely sculpts the distribution of asteroidal inclinations about $\nu_{6}$.  Therefore, we do not utilize criterion \textbf{E} in our current investigation.  Similarly, criterion \textbf{F} assessed the bulk volatile content of Earth analogs in Paper I and Paper II.  As our contemporary systems derive initial conditions by sampling planetesimal populations from \citet{walsh19} and \citet{clement20_psj}, the connection between objects accreted by Earth analogs and their original formation locations within the primordial nebula is not clearly defined.  Therefore, we do not consider criterion \textbf{F} in our present manuscript.

The low orbital eccentricities and inclinations of Earth and Venus in the modern solar system also represent key constraints for our simulations.  The dynamical excitation of a system of planets is typically quantified by its' angular momentum deficit \citep[AMD:][]{laskar97,chambers01}; a measure of a system's deviation from a perfectly circular, co-planar collection of orbits:

\begin{equation}
	AMD = \frac{\sum_{i}m_{i}\sqrt{a_{i}}[1 - \sqrt{(1 - e_{i}^2)}\cos{i_{i}}]} {\sum_{i}m_{i}\sqrt{a_{i}}} 
	\label{eqn:amd}
\end{equation}

As in Paper I and Paper II, we require our terrestrial systems (all planets with $m>$ 0.05 $M_{\oplus}$) finish with AMD $<$ 2AMD$_{SS}$ \citep[criterion \textbf{G}:][]{ray09a}; where AMD$_{SS}$ is the modern statistic for the four terrestrial planets (0.0018).

\section{Results}

\begin{figure*}
	\centering
	\includegraphics[width=.8\textwidth]{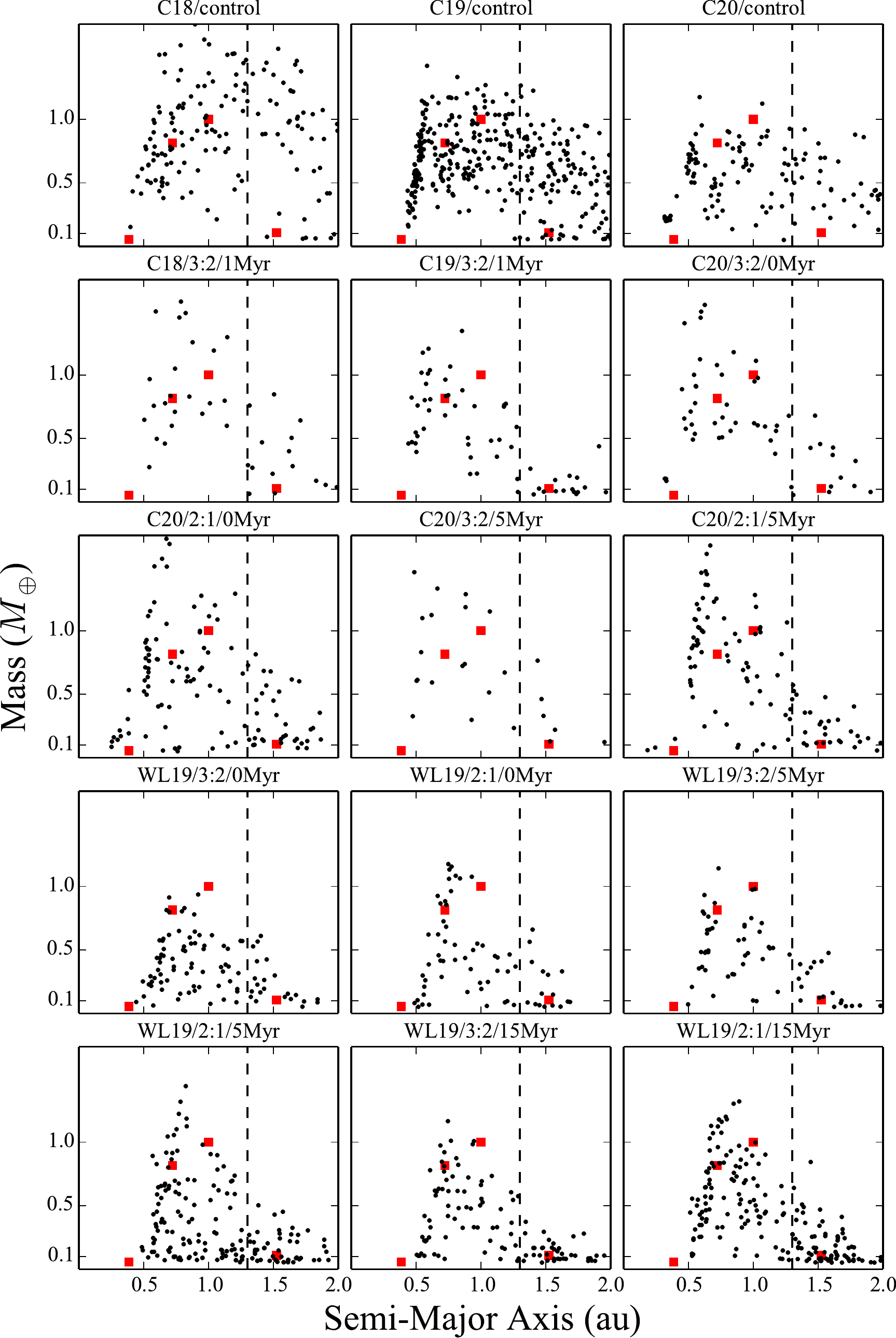}
	\caption{Distribution of semi-major axes and masses for all planets formed in all simulation sets.  The red squares denote the actual solar system values for Mercury, Venus, Earth and Mars.  The vertical dashed line separates the Earth and Venus analogs (left side of the line) and the Mars analogs (right side).  Details on the two instability models are tabulated in table \ref{table:gp} and the initial disk conditions for each set are summarized in table \ref{table:ics2}.}
	\label{fig:total}
\end{figure*}

\subsection{Comparison with Past Work}

The major conclusion of our investigation is that, to first order, our new instability simulations (C20 and WL19) produce qualitatively similar outcomes to the models of Paper I and Paper II.  This is particularly encouraging given that, in the absence of a giant planet instability model, the WL19 and C20 disks (that presumably represent more authentic initial conditions for the giant impact phase) produce too many planets, overly massive Mars analogs, and undersized Earth and Venus analogs.   Figure \ref{fig:total} plots the final masses and semi-major axes of all planets formed in each of our different simulation batches (table \ref{table:ics2}).  Indeed, the Mars analog masses (all planets with $a>$ 1.3 au) in all of our instability batches are consistently less than those in our three control sets.  Moreover, no Mars analog in an instability simulation finishes with $M>$ 1.0 $M_{\oplus}$ (we elaborate further on the instability's tendency to stunt Mars' growth in section \ref{sect:mars}).

It is clearly apparent from these plotted distributions that the WL19 disks systematically struggle to form sufficiently massive Earth and Venus analogs.  These differences are easily understood by inspecting the initial mass profile in each disk model.  The WL19 disks are comprised of 3.32 $M_{\oplus}$ of planet forming material between 0.7 and 3.0 au, with initial planetesimal semi-major axes selected such that the disk surface density profile falls off proportional to $r^{-3/2}$.  Thus, these disks nominally contain $\sim$1.13 $M_{\oplus}$ of terrestrial-forming material with $a<$ 1.3 au at time zero.  In order to form adequate Earth and Venus analogs and satisfy criteria \textbf{A} and \textbf{A1}, the respective analogs' feeding zones must stretch well into the Mars-forming region.  While it is common for Earth-analogs in classic N-body studies of terrestrial planet formation to accrete a significant fraction of objects from the outer regions of the disk \citep{ray07,fischer14,kaibcowan15}, the instability's tendency to void this region of large embryos conspires to severely limit the availability of such material in our WL19 instability runs.  Thus, while these simulations are successful in terms of suppressing the distribution of masses outside of $\sim$1.3 au, the analog systems themselves tend to provide poor matches to the actual system of terrestrial planets in terms of the masses of Earth and Venus (the majority of $M>$ 1.0 $M_{\oplus}$ planets occur in simulations that only form a single planet with $a<$ 1.3 au).  

The tendency of WL19 disks to form under-mass versions of Earth and Venus is slightly mitigated in our simulation batches investigating later instability delay times of 5 and 15 Myr.  Indeed, these batches are the only WL19 simulations in our sample to satisfy criteria \textbf{A} or \textbf{A1}, with the former (5 Myr) being more successful in terms of consistently yielding small Mars analogs \citep[i.e.: the instability occurs before Mars grows too large:][]{clement18,nesvorny21_tp}.  In such realizations, Earth and Venus are able to accrete a sufficient quantity of material from the outer disk prior to the instability reshaping the asteroid belt and Mars-forming region.  These results are best interpreted as favoring the instability's transpiration at an epoch where the terrestrial disk has attained an evolutionary state similar to our WL19/5Myr initial conditions; with roughly half the disk mass distributed in smaller planetesimals, and the remaining half locked in larger embryos (table \ref{table:ics2}).

In contrast to the WL19 models, our C20 disks begin with a total mass of terrestrial-forming material equal to 5.0 $M_{\oplus}$, distributed between 0.48 and 4.0 au with a surface density profile that is proportional to $r^{-3/2}$ (our reference C18 and C19 simulations essentially employ an identical disk in terms of its total mass and radial extent).  Therefore, our C20 simulations are initialized with $\sim$1.7 $M_{\oplus}$ of planet forming material in the Earth and Venus-forming regions ($a<$1.3 au); roughly equivalent to the modern cumulative mass of the two planets (1.815 $M_{\oplus}$).  

Figure \ref{fig:total} does not evince substantial differences between the distribution of planets formed in our C20/0Myr and C20/5Myr instability models.  This is not particularly surprising, given the fact that the ratio of total embryo to planetesimal mass does not meaningfully evolve between 0 and 5 Myr in the original C20 $GENGA$ simulations.  Instead, the largest difference between the two models is the actual masses of the larger embryos in the Earth and Venus-forming region (figure \ref{fig:kev_compare}).  Indeed, \citet{clement20_psj} noted that these embryo configurations emerge from the gas disk phase in a substantially-evolved, quasi-stable configuration that tends to hinder subsequent accretion events (note that our reference C20/control simulations finish the 200 Myr giant impact phase with 6.4 planets per system).  However, this is clearly not a problem in our C20 models that include a giant planet instability model as, to first order, the results of all four simulation batches (in terms of criterion \textbf{A1} and the distributions depicted in figure \ref{fig:total}) are largely consistent with those of the C18 and C19 reference instability simulations.  The largest outlier of our four C20 instability batches in terms of criterion \textbf{A} and \textbf{A1} is the C20/3:2/5Myr set, however we assess these discrepancies to largely be the consequence of small number statistics (the set only yields 20 total systems), rather than a systematic shortcoming of the 3:2 instability model.  Thus, our results indicate that a smaller number of more massive embryos in the Earth and Venus-forming regions around the time of nebular gas dissipation is a viable initial configuration for the early instability scenario.  While the cosmochemical implications of this scenario remain to be explored (e.g.: the consequences for differentiation, volatile delivery, volatile retention, etc.), bridging the gap between the runaway growth simulations of \citet{walsh19} and \citet{clement20_psj} and the modern terrestrial system is an encouraging finding of our present investigation.   We explore the causes of lower order variances in outcomes in the subsequent sections.

\begin{table*}
\centering
\begin{tabular}{c c c c c c c}
\hline
Set & A & A1 & B & C & D & G\\ 
& $a,m_{TP}$ & $m_{TP}$ & $\tau_{Mars}$ & $\tau_{\oplus}$ & $M_{AB}$ & AMD\\
\hline
C18/control & 0 & 0 & 9 & 86 & 2 & 8 \\
C18/3:2/1Myr & 26 & 26 & 12 & 95 & 20 & 14 \\
C19/control  & 2 & 2 & 33 & 80 & 2 & 15  \\
C19/3:2/1Myr & 35 & 35 & 23 & 80 & 70 & 12  \\
C20/control & 0 & 0 & 0 & 57 & 5 & 75\\
\hline
C20/3:2/0Myr & 12 & 20 & 45 & 28 & 72 & 15 \\
C20/2:1/0Myr & 8 & 20 & 39 & 37 & 66 & 17 \\
C20/3:2/5Myr & 0 & 5 & 50 & 33 & 95 & 0 \\
C20/2:1/5Myr & 11 & 27 & 45 & 35 & 75 & 12 \\
WL19/3:2/0Myr & 0 & 0 & 5 & 60 & 89 & 48 \\
WL19/2:1/0Myr & 0 & 0 & 7 & 0 & 86 & 40 \\
WL19/3:2/5Myr & 4 & 9 & 23 & 57 & 77 & 23 \\
WL19/2:1/5Myr & 4 & 6 & 42 & 54 & 88 & 19 \\
WL19/3:2/15Myr & 0 & 2 & N/A & 37 & 68 & 3 \\
WL19/2:1/15Myr & 0 & 0 & N/A & 27 & 82 & 6 \\
\hline	
\hline
\end{tabular}
\caption{Summary of percentages of systems which meet the various terrestrial planet success criteria established in table \ref{table:crit} (reference simulation statistics are reproduced from: \citet[][Paper I]{clement18} (C18), \citet[][Paper II]{clement18_frag} (C19), and \citet{clement20_psj} (C20)). The subscripts TP and AB indicate the terrestrial planets and asteroid belt respectively.}
\label{table:results}
\end{table*}

\subsubsection{Stunting Mars' growth}
\label{sect:mars}

\begin{figure}
	\centering
	\includegraphics[width=.5\textwidth]{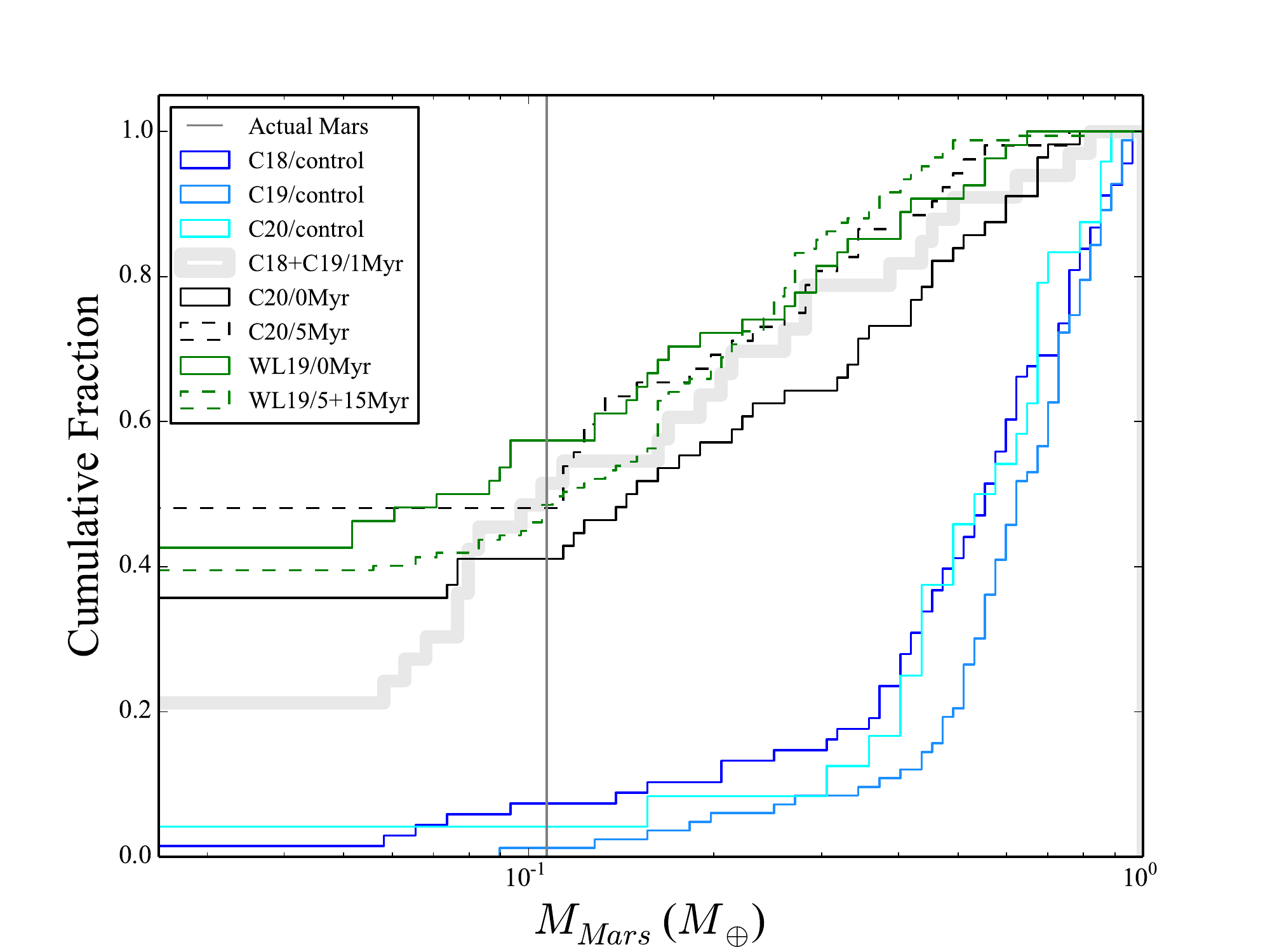}
	\caption{Cumulative distribution of Mars analog masses formed in our various simulation sets from this manuscript (black and green lines) compared with reference (see table \ref{table:ics2}) simulation batches from \citet{clement18,clement18_frag,clement20_psj}. Systems plotted in shades of blue represent simulations that did not incorporate a giant planet instability model from each respective past work.  The transparent thick grey line combines the 1 Myr instability delay reference simulations: C18/3:2/1Myr and C19/3:2/1Myr.  Black lines correspond to C20 disk simulations from this work, and green lines denote WL19 disk runs.  The solid and dashed lines for each data set separate 0 Myr instability delay simulations (solid) from the 5 and 15 Myr (dashed) batches.  The grey vertical lines corresponds to Mars’ actual mass.  Note that some systems may form multiple planets in this region, but here we only plot the most massive planet.  Systems that do not form a Mars analog via embryo accretion (i.e.: $M_{Mars}<$ 0.05 $M_{\oplus}$) are plotted as having zero mass.}
	\label{fig:mars}
\end{figure}

Our simulations confirm the results of Paper I and Paper II and broadly evidence the Nice Model instability's capacity to substantially restrict Mars' growth.  Figure \ref{fig:mars} plots the cumulative distribution of Mars analog masses in the various simulation sets from this study, and several reference models from our previous work (table \ref{table:ics2}). The similarity between the three blue curves (control runs without an instability), in addition to the consistency of the instability curves (green, black and grey lines) clearly demonstrates the transparency of Mars' formation to the particular terrestrial disk utilized (C18, WL19 or C20) and instability model employed (2:1 or 3:2).  It should be noted here that, given the resolution of our simulations, we intentionally plot all Mars analogs with $M<$ 0.05 $M_{\oplus}$ as possessing no mass.

The largest difference between the five instability curves in figure \ref{fig:mars} is the over-abundance of $\sim$0.05-0.08 $M_{\oplus}$ Mars analogs in the C18/3:2/1Myr and C19/3:2/1Myr reference simulations compared to our contemporary instability models.  We attribute these variances to the relative absence of embryos in this mass range within the Mars-forming region at the time of the instability in the WL19 and C20 disks (figure \ref{fig:kev_compare}).  As the largest embryos in the Mars-forming region of these disks already possess masses close to that of the real planet, it makes sense that few simulations yield smaller versions of the planet.

Our new instability simulations possess systematically improved success rates for criterion \textbf{D} (i.e.: they do not strand embryos in the asteroid belt) compared to the reference instability simulations from Paper I and Paper II.  There are two important factors responsible for these differences.  First, our new results are derived from an increased fraction of systems with adequately excited values of $e_{J}$ compared to our previous simulation batches (by virtue of our numerical pipeline: section \ref{sect:meth:restart}).  \citet{clement18_ab} demonstrated a positive correlation between $e_{J}$ and total depletion in the asteroid belt.  It therefore follows that $\gtrsim$70$\%$ of the runs in each of our simulation batches completely erode the asteroid belt region of larger embryos. Second, our simulations are initialized with a small number embryos in the asteroid belt.  Our WL19 models only possess a few embryos with semi-major axes just inside of 2.0 au (see figure \ref{fig:kev_compare}), and our C20 disks contain a system of five, $\sim$0.02 $M_{\oplus}$ embryos in the belt.  This is the consequence of the time required to reach the analytical isolation mass via runaway growth in the asteroid belt being much longer than the nebular gas dissipation timescale \citep{koko_ida_96}.  Conversely, the reference C18 and C19 simulations begin with $\sim$20, 0.025 $M_{\oplus}$ embryos in between 2.0 and 4.0 au.  Thus, particles in the asteroid belts of the WL19 $LIPAD$ and C20 $GENGA$ simulations simply are not permitted sufficient time to collisionally assemble and form an evolved, bimodal population of embryos and planetesimals.  As a result of utilizing these more realistic initial conditions in our contemporary study, we only note 15 examples of planets more massive than Mars forming in the asteroid belt within our entire batch of 356 instability simulations.

\begin{figure}
\centering
\includegraphics[width=.50\textwidth]{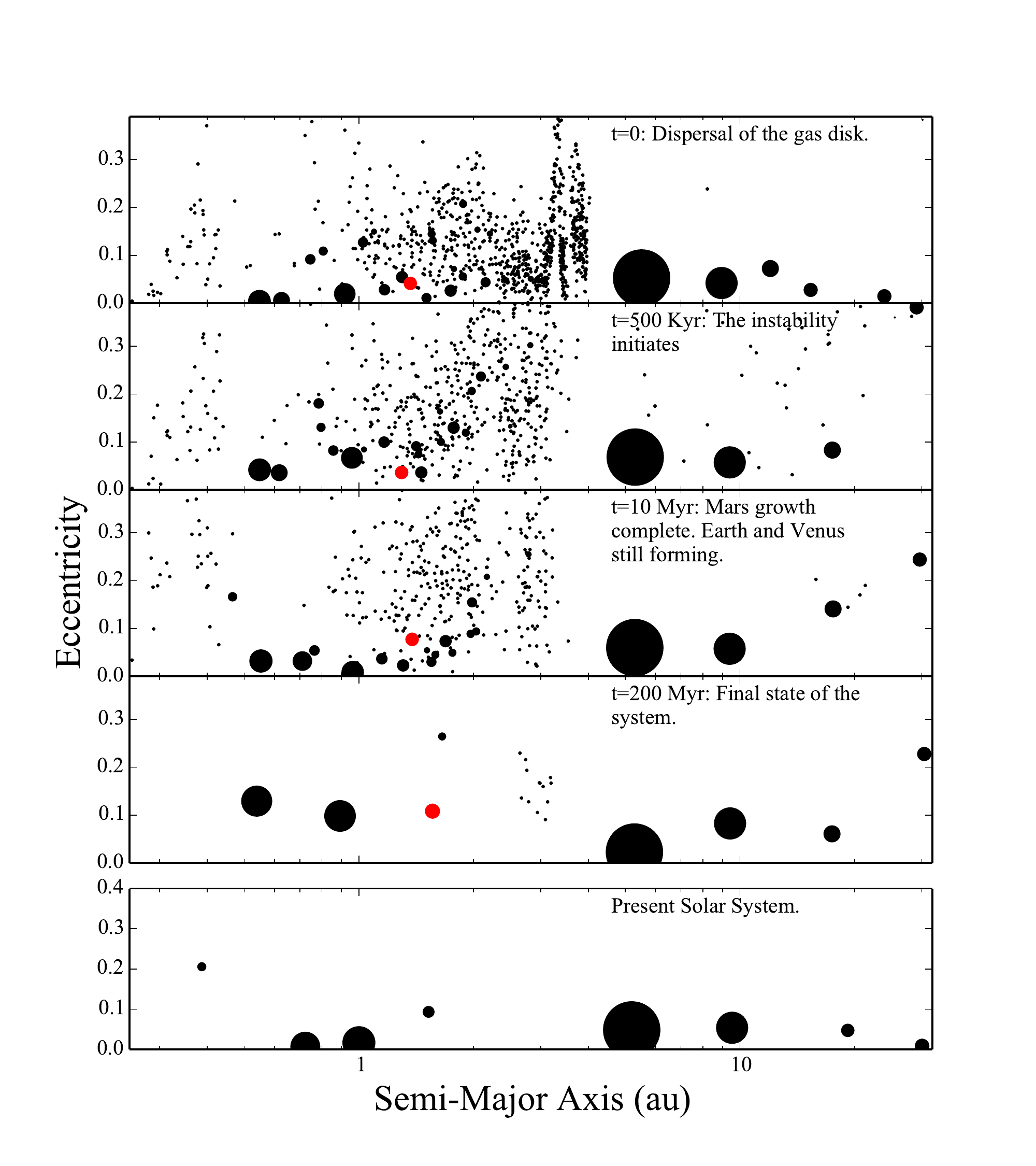}
\caption{Semi-Major Axis/Eccentricity plot depicting the evolution of a successful system in the C20/2:1/5Myr batch.  The size of each point corresponds to the mass of the particle (because Jupiter and Saturn are hundreds of times more massive than the terrestrial planets, we use separate mass scales for the inner and outer planets).  For reference, the embryo that becomes the Mars analog is plotted in red.  The final terrestrial planet masses are $M_{Venus}=$ 0.88 ($M_{Venus,SS}=$ 0.815), $M_{Earth}=$ 0.92 and $M_{Mars}=$ 0.18 ($M_{Mars,SS}=$ 0.107) $M_{\oplus}$.  The additional surviving embryo in the Mars region has a mass of 0.04 $M_{\oplus}$ (we extended this simulation and found that the embryo collided with the Sun at $t=$ 285 Myr).  This simulation also satisfies the four success criteria for the giant planets ($e_{55}=$ 0.023, $P_{S}/P_{J}=$ 2.36; compared with $e_{55,ss}=$ 0.044, $P_{S}/P_{J,ss}=$ 2.49)  described in \citet{nesvorny12} and \citet{clement21_instb}.}
\label{fig:time_lapse}
\end{figure}

More massive Mars analogs loosely correlate with shorter instability delay times (i.e. 0 Myr vs. 5 and 15 Myr) for both the C20 and WL19 simulation batches depicted in figure \ref{fig:mars}.  This result is also evident in the distributions of planets plotted in figure \ref{fig:total}.  Similarly, our C20/0Myr simulations possess lower success rates for criterion \textbf{D} than the C20/5Myr models.  Paper I analyzed these trends in detail and concluded that instabilities ensuing precipitously from less-processed terrestrial disks were less successful as the disk tends to re-spread after being truncated by the instability through interactions between embryos and overly-abundant planetesimals.  When this is the case, fully evolved systems often contain 3-4 similarly massed planets.  Thus, Mars analogs tend to be too massive, while Earth and Venus are too small.  However, both the WL19 and C20 disks tested in our current study attain advanced evolutionary states around the time of nebular gas dissipation.  As a result, these trends are not as pronounced as in Paper I and Paper II.  Specifically, this is a result of the fact that the embryo-planetesimal mass partitioning in the Mars-forming region is similar in all of our simulation sets (roughly half of the mass in embryos and half in planetesimals).  Thus, we conclude that all three timings (0, 5 and 15 Myr) investigated in our current work are viable in terms of consistently replicating Mars' mass.  However, as discussed in the previous section, constraints related to Earth and Venus' mass lead us to disfavor delays of 0 and 15 Myr for the WL19 disks.  Moreover, we do not test delays longer than 5 Myr with our C20 disks as embryos in the Mars region grow too massive after $t\simeq$ 5 Myr.  Figure \ref{fig:time_lapse} plots an example successful evolution from our preferred C20/2:1/5Myr batch of simulations.

\subsubsection{Angular Momentum Deficit}
\label{sect:amd}

\begin{figure}
	\centering
	\includegraphics[width=.5\textwidth]{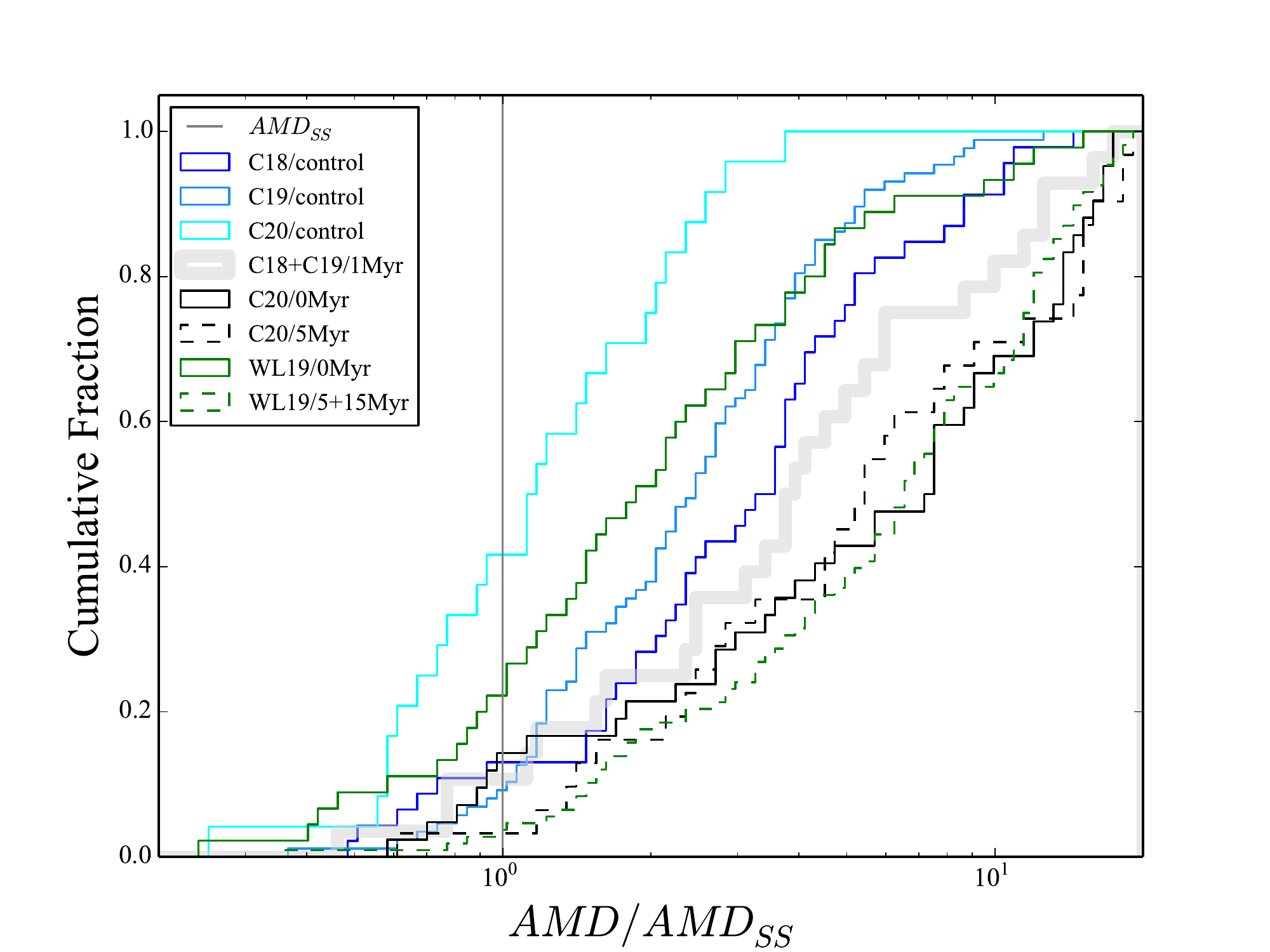}
	\caption{Cumulative distribution of AMDs (normalized to the solar system value for the four terrestrial planets: equation \ref{eqn:amd}) formed in the various simulation sets from this manuscript (black and green lines) compared with reference simulation batches from \citet[][see table \ref{table:ics2}]{clement18,clement18_frag,clement20_psj}. Systems plotted in shades of blue represent simulations that did not incorporate a giant planet instability model.  The transparent thick grey line combines the 1 Myr instability delay reference simulations C18/3:2/1Myr and C19/3:2/1Myr.  Black lines correspond to C20 disk instability simulations from this work, and green lines denote WL19 disk runs.  The solid and dashed lines for each data set separate 0 Myr instability delay simulations (solid) from the 5 and 15 Myr (dashed) batches.}
	\label{fig:amd}
\end{figure}

Figure \ref{fig:time_lapse} highlights the challenge of replicating the low orbital eccentricities and inclinations of Earth and Venus in instability models that adequately excite the orbits of the gas giants.  Indeed, the Earth and Venus analogs in this simulation each possess $e\simeq$ 0.10; substantially more dynamically excited than the nearly circular orbits they inhabit in the actual solar system.  Figure \ref{fig:amd} plots the cumulative distribution of AMDs in our various simulation sets, compared with those of systems formed in our reference simulations from Paper I, Paper II and \citet{clement20_psj}.  In general, our instability simulations yield a fraction of systems with AMDs less than twice the solar system value (table \ref{table:results}) that is similar to those formed in control disks with an instability model (though the median values are clearly different: blue lines in figure \ref{fig:amd}).  Moreover, we note no significant differences in the final AMDs produced in simulations invoking different giant planet instability models (2:1 and 3:2; see table \ref{table:gp}).  This demonstrates that perturbations from the strong $\nu_{5}$ resonance being initialized closer to the Earth-forming region in our 2:1 models do not adversely affect the final distribution of AMDs.

  While the angular momentum deficit of the terrestrial system is still a problem in our new models, our contemporary results are no worse than those of our previous models.  As our simulation pipeline (section \ref{sect:meth:restart}) is specifically designed to select for stronger instability that yield higher gas giant eccentricities, one might naively expect our new instabilities to systematically possess larger AMDs than the C18/1Myr and C19/1Myr reference models.  However, our contemporary integrations actually perform about the same or, in certain cases, slightly better when measured against criterion \textbf{G}.  Indeed, all four 0 Myr delay simulation batches boast higher success rates for \textbf{G} and marginally-improved distributions of AMDs in figure \ref{fig:amd}.  It is therefore likely that this improvement (or lack of retrogression) is related to the structure of the WL19 and C20 disks themselves.  This is supported by the fact that 75$\%$ of our C20/control simulations satisfy criterion \textbf{G}.  Embryos in the Earth and Venus-forming region of the WL19 and C20 disks attain relatively large masses (figure \ref{fig:kev_compare}) while still engulfed in the dense nebular gas.  As a result, the embryos emerge from the gas disk on dynamically cold orbits, and require only a few additional large accretion events to reach their final masses (section \ref{sect:tgrow}).  As these embryos experience fewer large encounters that might excite their eccentricities in these more processed terrestrial disks, they tend to survive the planet formation process with lower eccentricities and inclinations than the Earth and Venus analogs in our C18 and C19 reference models.

While the solar system outcome lies comfortably within the lower range of AMDs generated in all of our various instability simulations (green and black lines of figure \ref{fig:amd}), this result is potentially misleading since overly-massive Mars analogs with low eccentricities can positively modify the AMDs of systems with overly-excited Venus and Earth analogs towards successful values.  Given the preponderance of systems that resemble the one plotted in figure \ref{fig:time_lapse}, we conclude that the replication the low AMD of the modern terrestrial system represents a significant outstanding problem for future models to resolve.  In addition to focusing on the particularly problematic eccentricities and inclinations of Earth and Venus \citep[rather than the entire terrestrial system:][]{nesvorny21_tp}, such investigations should comprehensively attempt to account for imperfect collisions \citep{chambers13,clement18_frag,deienno19} and the effects of the dissipating gas disk \citep[e.g.:][]{morishma10,matsu10,levison15,carter15}.

\subsubsection{Fewer Giant Impacts}
\label{sect:tgrow}

\begin{figure*}
	\centering
	\includegraphics[width=.49\textwidth]{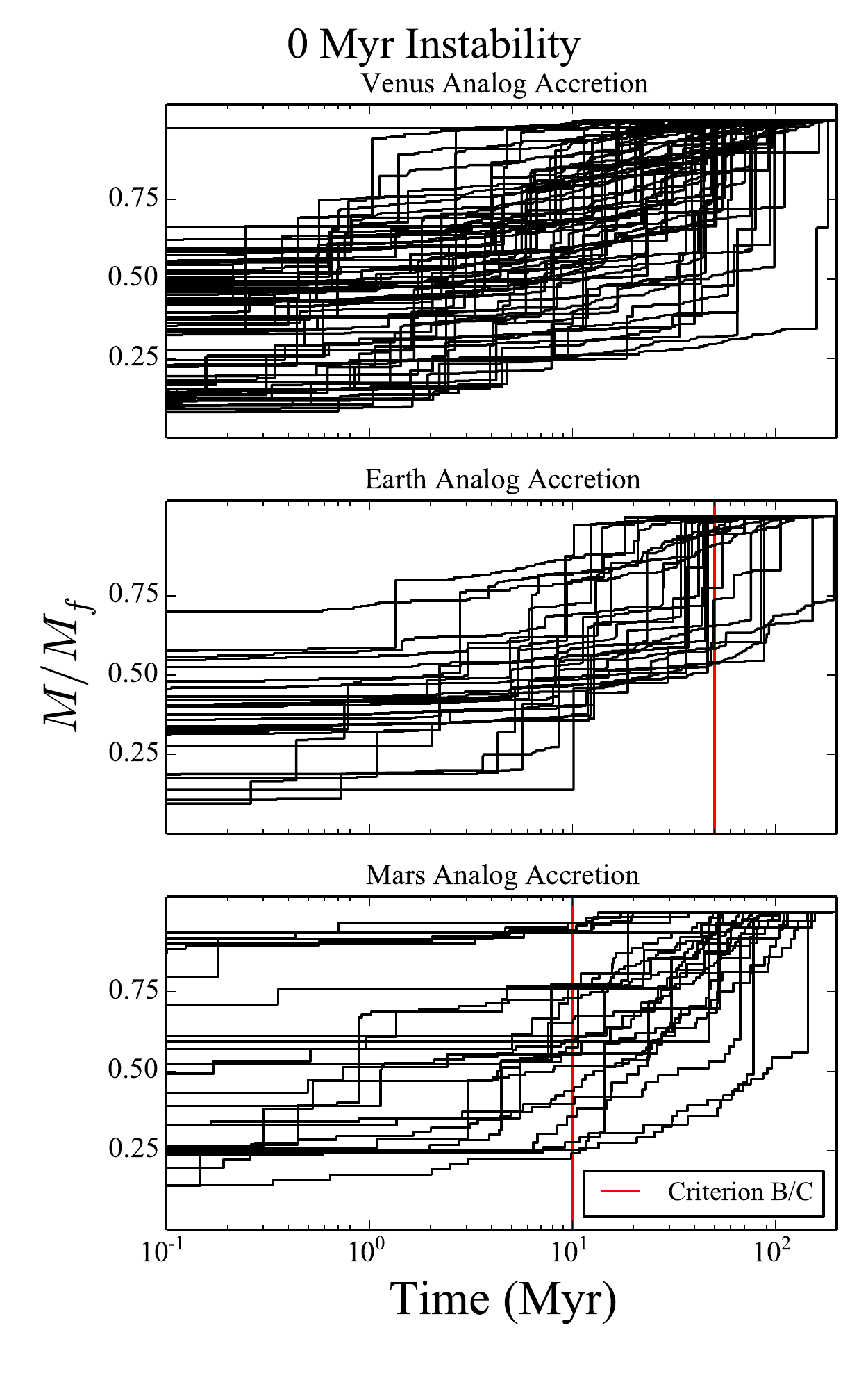}
	\includegraphics[width=.49\textwidth]{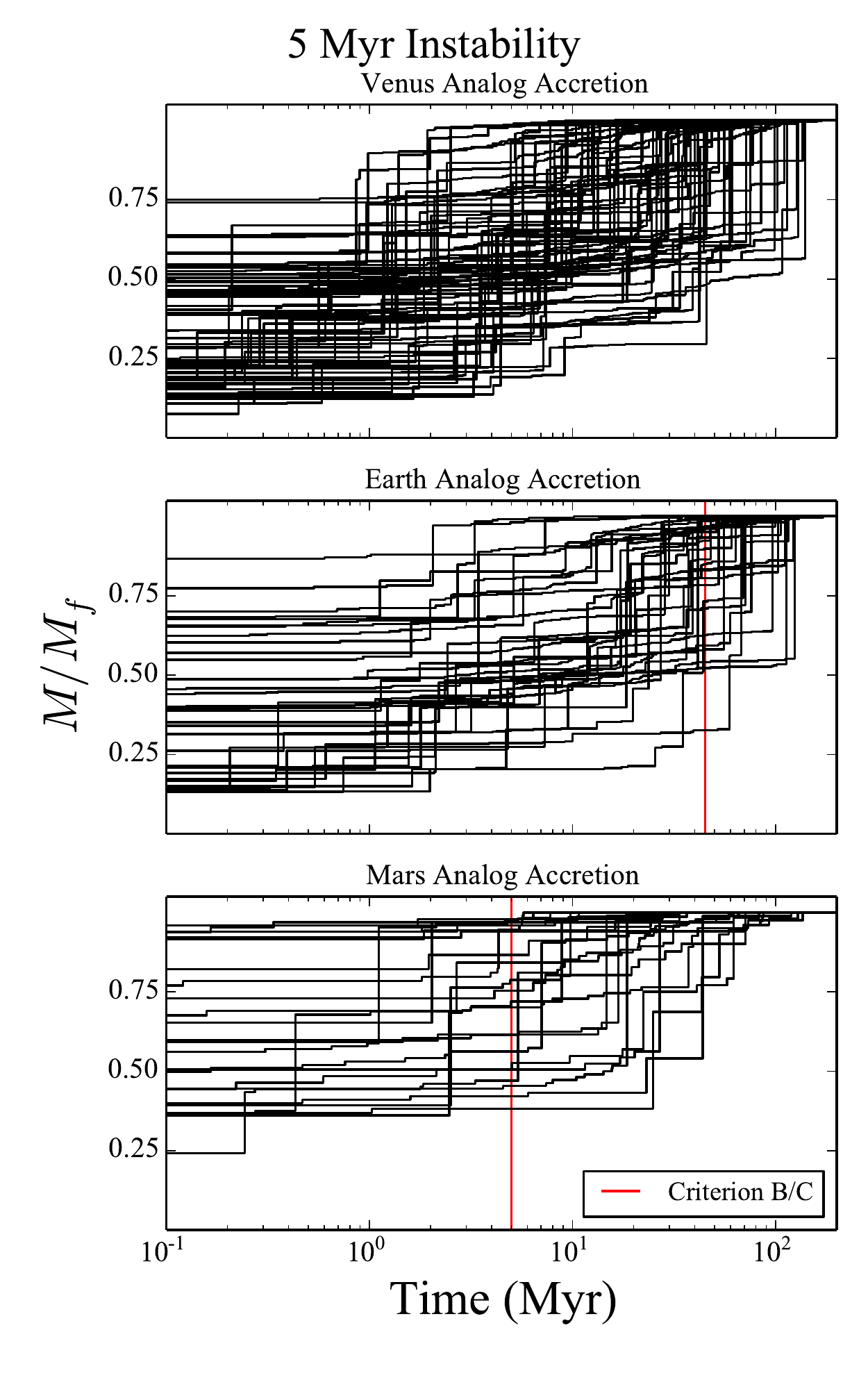}
	\caption{Accretion histories for analog planets in our various simulation sets investigating 0 and 5 Myr instability delays (left and right panels, respectively, note that time zero correlates with the beginning of the simulation).  Venus analogs (top panels) correspond to all planets with $M>$ 0.6 $M_{\oplus}$ and $a<$ 0.85 au; Earth analogs consists of planets with $M>$ 0.6 $M_{\oplus}$ and 0.85 $<a<$ 1.3 au, while Mars analogs are defined as all objects with 0.05 $<M<$ 0.3 $M_{\oplus}$ and 1.3 $<a<$ 2.0 au.  Each planet's mass is normalized to its final value and plotted as a function of time over the duration of the simulation.  The red vertical lines correlate with our success criteria \textbf{B} ($<$10 Myr with respect to time zero) and \textbf{C} ($>$50 Myr) for the respective growth timescales of Mars and Earth analogs.}
	\label{fig:tgrow}		
\end{figure*}

Earth analogs in our simulations typically reach their final masses within the first $\sim$50 Myr following the instability's onset via a final giant impact with a roughly equal-mass impactor.  This is not particularly surprising given the more advanced evolutionary states of the Earth and Venus-forming regions in our WL19 and C20 disks at the beginning of our simulations.  In all of our simulation batches, this regime is dominated by 3-5, $\sim$0.1-0.4 $M_{\oplus}$ embryos around the time of the Nice Model's onset.  As a consequence of increased eccentricity excitation in the inner solar system after the instability, these large proto-planets precipitously attain crossing orbits that lead then to quickly combine and form Earth and Venus-analogs.  Figure \ref{fig:tgrow} depicts the accretion histories for all Earth ($M>$ 0.6 $M_{\oplus}$; 0.85 $<a<$ 1.3 au), Venus ($M>$ 0.6 $M_{\oplus}$; $a<$ 0.85 au) and Mars (0.05 $<M<$ 0.3 $M_{\oplus}$; 1.3 $<a<$ 2.0 au) analog planets formed in our various simulation sets.  Table \ref{table:tgrow} summarizes the same data, and provides the percentages of analogs accreting 90$\%$ of their ultimate masses within the first 10, 50 and 100 Myr following the instability for each of our simulation batches investigating different instability delays.  Though the total fraction of Earth analogs with growth timescales in excess of 50 Myr (criterion \textbf{C}) in each of our simulation batches is less than for the C18/3:2/1Myr and C19/3:2/1Myr reference sets, a reasonable fraction of systems still experience late final giant impacts that are consistent with the inferred timing of the Moon-forming impact \citep{kleine09,rudge10,kleine17}.  

We note that the prevalence of equal-mass final impacts on Earth is potentially consistent with the Moon-formation scenario envisioned in \citet{canup12}.  While these conditions are fairly improbable in the classic terrestrial planet formation model \citep[a $\sim$Mars-mass impactor is far more common:][]{chambers98,kaibcowan15}, they occur fairly regularly when both our WL19 and C20 disks are perturbed by the giant planet instability.  

In contrast to the formation of Earth, many of our Venus analogs form within a few tens of Myr without experiencing a significant final giant impact.  This result is consistent with the dichotomous histories of Earth and Venus proposed in \citet{jacobson17b} to reconcile Venus' lack of an internally generated dynamo and natural satellite.  Additionally, figure \ref{fig:tgrow} and table \ref{table:tgrow} clearly demonstrates the tendency of Mars analogs in our 0 Myr instability delay batches to experience prolonged accretion histories (criterion \textbf{B}) that are inconsistent with the planets' inferred geological formation timescale \citep{Dauphas11,kruijer17_mars}.  This is a result of the disk re-spreading after the instability (discussed in section \ref{sect:mars}).  Conversely, the accretion histories of our 5 Myr instability delay Mars analogs (right panel of figure \ref{fig:tgrow}) and their corresponding success rates for criterion \textbf{B} provide a better match to the actual planets' presumed rapid growth (an improvement from our reference C18/3:2/1Myr and C19/3:2/1Myr instability sets from Papers I and II).

\begin{table}
\centering
\begin{tabular}{c c c c}
\hline
$t_{inst}$ & $t_{90}<$ 10 Myr & $t_{90}<$ 50 Myr & $t_{90}<$ 100 Myr \\ 
\hline
\multicolumn{2}{c}{\textbf{0 Myr instabilities}} & & \\
\hline
Venus & 13 & 53 & 84 \\
Earth & 0 & 58 & 84 \\
Mars & 23 & 48 & 8 \\
\hline
\multicolumn{2}{c}{\textbf{5 Myr instabilities}} & & \\
\hline
Venus & 15 & 65 & 84 \\
Earth & 11 & 57 & 83 \\
Mars & 48 & 84 & 96 \\
\hline
\multicolumn{2}{c}{\textbf{15 Myr instabilities}} & & \\
\hline
Venus & 16 & 65 & 93 \\
Earth & 20 & 70 & 85 \\
Mars & 47 & 84 & 94 \\
\hline
\end{tabular}
\caption{Summary of percentages of planet analogs accreting 90$\%$ of their ultimate mass within the first 10, 50 and 100 Myr (columns 2-4) after the instability.  Note that, in contrast to our calculations for criterion \textbf{B} and \textbf{C} (the accretion time with respect to time zero: table \ref{table:crit}), here we compute the growth timescale with respect to the instability's onset.}
\label{table:tgrow}
\end{table}

\subsection{Jupiter's eccentricity excitation}
\label{sect:e55}

\begin{figure}
    \centering
	\includegraphics[width=.5\textwidth]{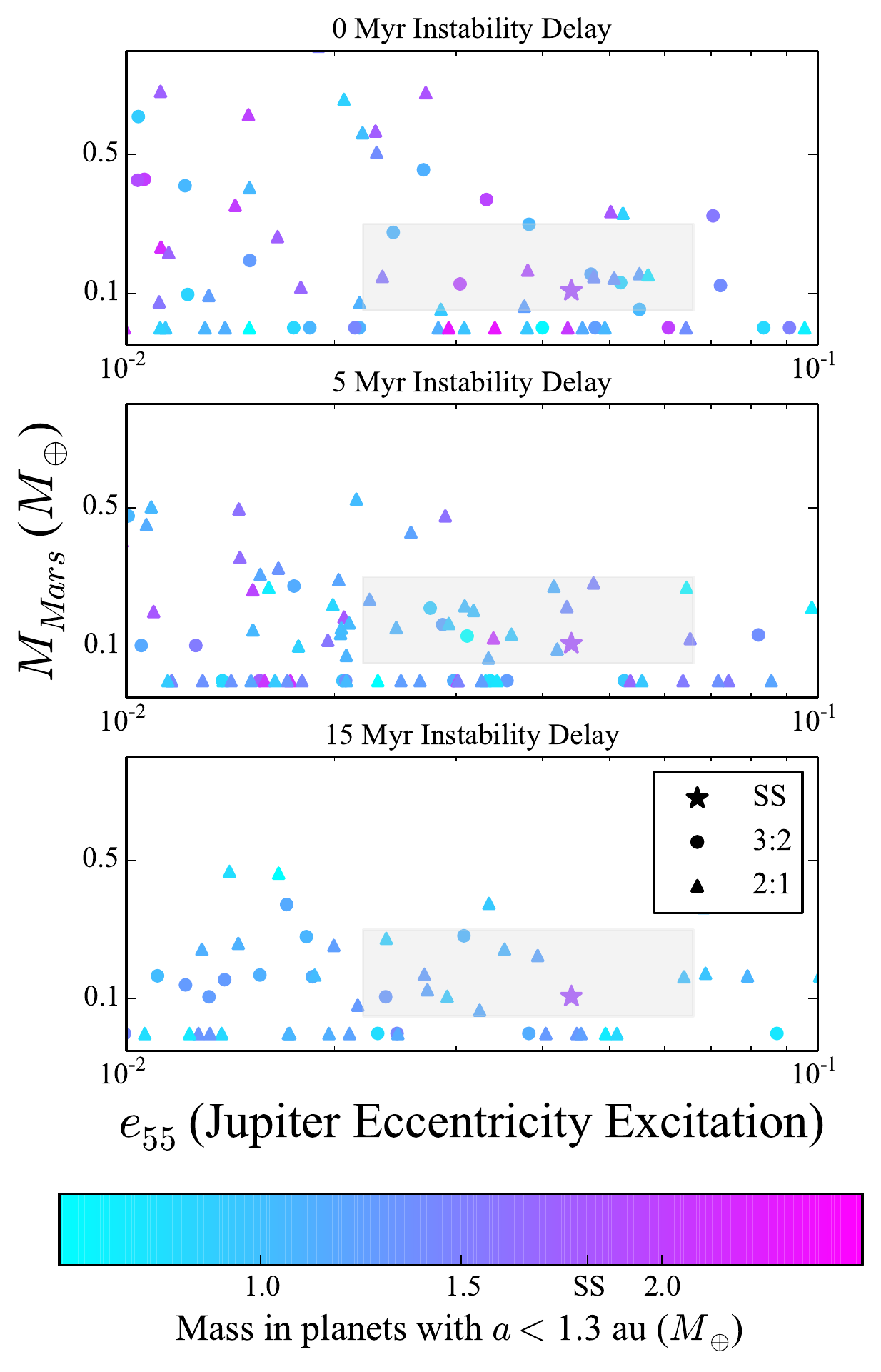}
	\caption{Dependence of depletion in the inner solar system on Jupiter's eccentricity excitation (quantified here by $e_{55}$: the magnitude of the solar system's fifth eccentric eigenfrequency, $g_{5}$, in Jupiter's eccentricity).  Only simulations with $e_{56}<e_{55}$ are plotted (i.e. those that do not over-excite Saturn).  Each sub-panel displays results from our various batches of simulations investigating different instability delay times (0, 5 and 15 Myr: table \ref{table:ics2}).  Each point plots Mars' final mass against $e_{55}$ (Mars analogs with $M<$ 0.05 $M_{\oplus}$ are plotted as possessing no mass).  The color of each point corresponds to the total remaining planetary mass with $a<$ 1.3 au (nominally Earth and Venus analogs).  Simulations utilizing a 3:2 Jupiter-Saturn resonance instability model are plotted with circular points, and 2:1 models are assigned triangular markers.  The stars plot the solar system values of $M_{Mars}$ and $e_{55}$.  The grey shaded regions encompass simulations with 0.05 $<M_{Mars}<$ 0.3 $M_{\oplus}$ (criterion \textbf{A}) and $e_{55}$ within 50$\%$ of the solar system value \citep{clement21_instb}.}
	\label{fig:e55}
\end{figure}

Our simulations indicate that the instability's tendency to sufficiently stunt Mars' growth correlates with the adequate excitation of Jupiter's eccentricity towards the solar system value.  Figure \ref{fig:e55} plots the relationship between Mars' mass and the excitation of Jupiter's fifth eccentric mode ($e_{55,SS}=$ 0.044; closely related to the planets' mean eccentricity in the modern solar system).  To determine $e_{55}$, we integrate each remnant system of planets for an additional 10 Myr and compute the secular frequencies and relative coefficients for each planet via frequency modified Fourier transform \citep{nesvorny96}.  We also analyzed the connection between $M_{Mars}$ and the other eccentric modes of the Jupiter-Saturn system and concluded that depletion in the outer terrestrial disk correlates most strongly with $e_{55}$.  As many of our instabilities over-excite Saturn's eccentricity (quantified here by its eccentric forcing on Jupiter: $e_{56,SS}=$ 0.0157), we only display systems that finish in the $e_{55}>e_{56}$ regime \citep[e.g.:][]{clement21_instb}.  By substantially increasing our sample of systems with $P_{S}/P_{J}<$ 2.5, high $e_{55}$, and low $e_{56}$, our results reinforce the general trends observed in Paper I.  Outcomes where $M_{Mars}\gtrsim$ 0.3 $M_{\oplus}$ (criterion \textbf{A}) are almost exclusively derived from simulations that do not properly excite $e_{55}$.  Exceptions to this trend are limited to WL19/0Myr runs where the instability occurs too early; thus causing the remnant terrestrial disk to spread and repopulate the Mars-forming region with embryos and planetesimals from the Earth-forming region (section \ref{sect:mars}).

While a statistical comparison between our 2:1 and 3:2 instability models (table \ref{table:gp}) is complicated by our simulation pipeline that restarts integrations that experience unideal jumps or excessive residual migration (section \ref{sect:meth:restart}), it is clear from figure \ref{fig:e55} that our new 2:1 scenarios yield similar results to the 3:2 cases typically considered in the literature.  To crudely control for the effects of our simulation pipeline, table \ref{table:e55_perrat} reports the percentage of simulations that effectively limit Mars' mass to less than that of the real planet from the subset of all realizations where $e_{55}$ is excited to at least $50\%$ of its modern value.  49$\%$ of all our 2:1 models are successful in this manner, as compared with 57$\%$ of our 3:2 runs.  If we increase our limiting value for Mars' mass to 0.3 $M_{\oplus}$ (criterion \textbf{A}) 87$\%$ of our 2:1 simulations and 92$\%$ of our 3:2 models adequately exciting $e_{55}$ possess a small Mars.  Thus, while the subset of simulations that reasonably replicate Jupiter's eccentricity are highly successful at limiting Mars' mass (regardless of giant planet instability model employed), our 3:2 sets tend to more efficiently restrict Mars' growth.

Figure \ref{fig:e55} demonstrates how Earth and Venus' ability to continue to accrete material after the instability and attain the correct masses is not hindered when $e_{55}$ is properly excited.  The color of each point in the plot depicts the total planetary mass in the Earth-Venus region ($a<$ 1.3 au).  Nominally this corresponds to the cumulative mass of Earth and Venus when criterion \textbf{A} is satisfied, however we include systems that form different numbers of planets (i.e.: one $\sim$ 2.0 $M_{\oplus}$ planet, or three smaller planets) to decouple the stochastic nature of the peculiar series of final impacts that generates the Earth-Venus system. The fact that multiple dark blue and purple points fall within the grey shaded regions demonstrates that the solar system result falls well within the spectrum of outcomes produced by our simulations.  While the low masses of Earth and Venus analogs in the 15 Myr instability delay simulations are related to features of the WL19 disks discussed in \ref{sect:mars}, the prevalence of light blue points throughout all panels of the figure exposes the fragility of the terrestrial system during the violent and stochastic Nice Model instability \citep[e.g.:][]{agnorlin12,kaibcham16}.  Indeed, many simulations yield a residual terrestrial system that bares little resemblance to the actual inner solar system if, for example, the gas giants spend lengthy periods of time on highly-eccentric orbits or the ejected ice giant makes an inordinate number of perihelia passages through the inner solar system.  Similarly, the number of small Mars analogs formed in simulations with excessively low final $e_{55}$ values evidences the stochasticity of the instability.  As our numerical pipeline (section \ref{sect:meth:restart}, point 3) necessarily selects for systems that attain high-$e_{J}$ values during the instability, these low-$M_{Mars}$/low-$e_{55}$ outcomes occur when the eccentricity excitation in the outer solar system is sufficient (i.e.: the gas giants attain high eccentricities for long enough) to significantly evacuate material from the Mars-forming region before the giant planets' orbits damp excessively during the residual migration phase \citep[e.g.:][]{nesvorny12}.

\citet{nesvorny21_tp} argued that depletion in the Mars and asteroid belt regions is fairly independent of the giant planets' peculiar evolution (i.e.: eccentricity excitation during the instability, depth and duration of encounters with the ejected planet, etc.).  Rather, the authors posited that the dominant mechanism responsible for limiting Mars' mass is the giant planets' attaining non-zero eccentricities before Mars exceeds its modern mass.  Thus, the early instability scenario is essentially an extension of the $EEJS$ (Extra Eccentric Jupiter and Saturn) model of \citet{ray09a}.  Regardless of whether the giant planets originate on eccentric orbits, or acquire them via an early Nice Model instability, the results are qualitatively the same.  Specifically, \citet{nesvorny21_e_nep} studied three independent instability evolutions (case 1 and 2 from \citet{nesvorny13} and the model of \citet{deienno18}).  However, these three instabilities investigated by the authors transpired at dissimilar epochs (6, 20 and 0.6 Myr, respectively), thus complicating an interpretation of the effects of each particular model.  Our contemporary simulations largely confirm the overall conclusions of \citet{nesvorny21_tp}.  The tendency of the instability to limit Mars' mass correlates most strongly with the adequate excitation of Jupiter's eccentricity.  As all simulations that sufficiently excite $e_{55}$ finish with a small Mars, it is clear that the early instability scenario is viable for a broad spectrum of plausible evolutionary pathways for the giant planets; provided the evolution culminates in a solar system-like giant planet architecture.

\begin{table*}
\centering
\begin{tabular}{c c c}
\hline
Constraint & 2:1 & 3:2 \\ 
\hline
$N(M_{Mars} \leq 0.107)$ / $N(e_{55} \geq 0.022)$ & 49 & 57 \\
$N(M_{Mars} \leq 0.107)$ / $N(P_{S}/P_{J} \leq 2.5)$ & 49 & 34 \\
$N(M_{Mars} \leq 0.107)$ / $N(e_{55} \geq 0.022$ $\&$ $P_{S}/P_{J} \leq 2.5)$ & 47 & 47 \\
\hline
\end{tabular}
\caption{Summary of percentages of systems from our simulation batches investigating different instability models (2:1 and 3:2; table \ref{table:gp}) simultaneously satisfying important constraints for the inner and outer solar system.  The rows are as follows: (1) percentage of all systems adequately exciting Jupiter's eccentricity ($e_{55}>$ 0.022) that produce a small Mars ($M_{Mars}<$ 0.3 $M_{\oplus}$), (2) percentage of systems with $P_{S}/P_{J}<$ 2.5 that possess a small Mars, and (3) small Mars-hosting systems from the subset of simulations simultaneously successful in terms $e_{55}$ and $P_{S}/P_{J}$.}
\label{table:e55_perrat}
\end{table*}

\subsection{Residual migration beyond Jupiter and Saturn's 5:2 MMR}
\label{sect:2.5_vs_2.8}

An unanswered question from our previous investigations of the early instability scenario in Paper I and Paper II centers around instability evolutions that achieved 2.5 $<P_{S}/P_{J}<$ 2.8.  Systems within this range were considered in both papers to attain adequate statistics as outcomes where the gas giants finish inside of their mutual 5:2 MMR are slightly uncommon in statistical studies of the instability (only $\sim$17$\%$ of both our 2:1 and 3:2 models are successful in this manner\footnote{Note that the rarity of attaining $P_{S}/P_{J}<$ 2.5 in and of itself is not a systematic issue in statistical studies of the Nice Model as the solar system result is merely one possibility within a broad spectrum of final values of $P_{S}/P_{J}$.  However, the tendency of $P_{S}/P_{J}<$ 2.5 instabilities to under-excite $e_{55}$ is a systematic issue in the sense that the solar system result, while possible, is a statistical outlier \citep[particularly for the 3:2 instability:][]{clement21_instb}.}).  As the dominant secular resonances currently in the inner asteroid belt presumably perturb the Mars-forming region more strongly in evolutions where $P_{S}/P_{J}$ exceeds the modern value, it is unclear whether the incorporation of these results in our previous work artificially boosted the fraction of systems finishing with a small Mars.  In the modern asteroid belt, the zero-inclination, eccentricity-averaged locations of the $\nu_{6}$ and $\nu_{16}$ resonances are at $\sim$ 2.05 and $\sim$ 1.95 au, respectively.

\begin{figure}
    \centering
	\includegraphics[width=.5\textwidth]{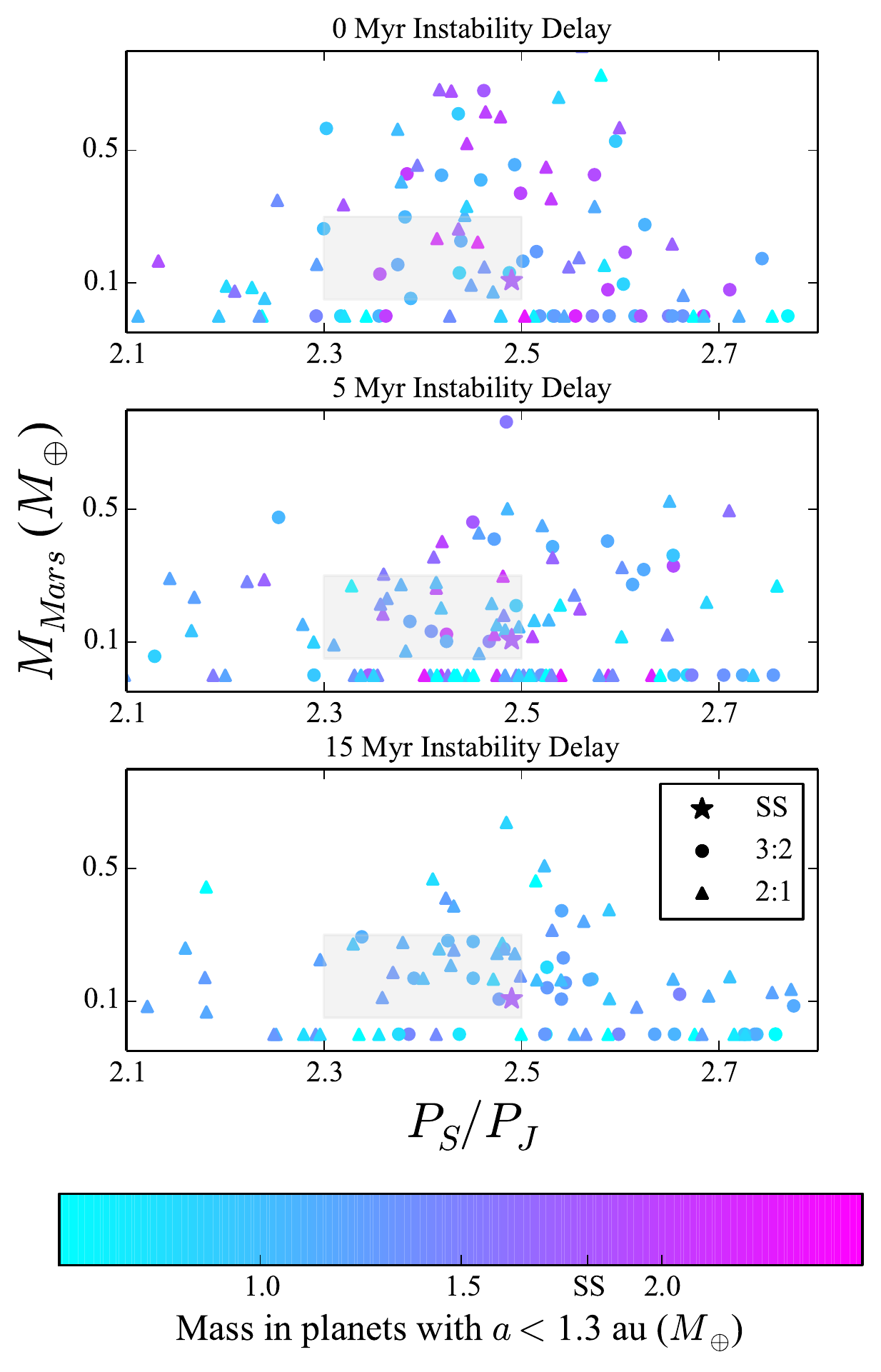}
	\caption{Similar to figure \ref{fig:e55}, except here each point plots the dependence of Mars' final mass on the ultimate Jupiter-Saturn period ratio (Mars analogs with $M<$ 0.05 $M_{\oplus}$ are plotted as possessing no mass).  The different sub-panels display simulation results from our various batches of simulations investigating different instability delay times (0, 5 and 15 Myr: table \ref{table:ics2}).    The color of each point corresponds to the total remaining mass in planets with $a<$ 1.3 au (nominally Earth and Venus analogs).  Simulations utilizing a 3:2 Jupiter-Saturn resonance instability model are plotted with circular points and 2:1 models are assigned triangular markers.  The stars plot the solar system values of $M_{Mars}$ and $P_{S}/P_{J}$.  The grey shaded regions encompass simulations with 0.05 $<M_{Mars}<$ 0.3 $M_{\oplus}$ (criterion \textbf{A}) and 2.3 $<P_{S}/P_{J}<$ 2.5 \citep{nesvorny12}.}
	\label{fig:perrat}
\end{figure}

Figure \ref{fig:perrat} plots the mass of Mars against the final value of $P_{S}/P_{J}$ in each of our various simulation sets (analogs with $M_{Mars}<$ 0.05 $M_{\oplus}$ and those forming no Mars analog are depicted as possessing zero mass).  As in figure \ref{fig:e55}, the color of each point corresponds to the total mass in planets with $a<$ 1.3 au.  In this manner, dark blue and purple points falling in the grey shaded region delimiting 2.3 $<P_{S}/P_{J}<$ 2.5 and 0.05 $<M_{Mars}<$ 0.3 should be interpreted as successful in terms of truncating the terrestrial disk in the Mars-forming region without limiting the accretion of the remaining inner planets.  It is clear from figure \ref{fig:perrat} that appropriate inner solar system mass distributions occur across the full range $P_{S}/P_{J}$ values produced in our simulations.

A loose trend of enhanced Mars and total Earth-Venus masses appears to manifest around 2.4 $\lesssim P_{S}/P_{J} \lesssim$ 2.6, near the solar system value of 2.49.  Thus, the largest final period ratios do tend to correlate with marginally depressed Mars masses, however this is partially a consequence of small number statistics and the fact that these values are unlikely outcomes of our simulation pipeline (section \ref{sect:meth:restart}).  While we cannot discount the fact that residual migration beyond Jupiter and Saturn's mutual 5:2 resonance likely artificially suppresses the masses of Mars analogs, figure \ref{fig:perrat} depicts numerous realizations that are exemplarily solar system analogs in terms of simultaneously matching the mass distribution in the terrestrial region and $P_{S}/P_{J}$.  As the solar system result is clearly well within the spectrum of outcomes produced in our simulations, we conclude that variances resulting from the stochastic nature of terrestrial planet formation and the Nice Model instability significantly outweigh the effects of residual migration past $P_{S}/P_{J}=$ 2.5. 

When we isolate systems finishing with $P_{S}/P_{J}$, as in section \ref{sect:e55} (table \ref{table:e55_perrat}), we find that our 2:1 models are slightly more successful at restricting Mars' growth than the 3:2 simulations (49$\%$ versus 34$\%$ of simulations limiting Mars' mass to less than its actual value).  These trends essentially counterbalance the results discussed in section \ref{sect:e55} for simulations adequately exciting $e_{55}$.  Thus, identical percentages (47$\%$) of both our 3:2 and 2:1 simulations simultaneously attaining $e_{55}>$ 0.022 and $P_{S}/P_{J}<$ 2.5 also produce a Mars that is less massive than the real planet (this statistic increases to 89$\%$ when we consider Mars analogs with $M_{Mars}<$ 0.3 $M_{\oplus}$).  Thus, we conclude that the primordial 2:1 Jupiter-Saturn resonance is an equivalently viable initial condition (compared to the 3:2) for the early instability scenario.

\subsection{Mercury Analogs}
\label{sect:merc}

In the previous sections we neglected the formation of Mercury.  Indeed, Mercury's low mass compared to the other inner planets ($M_{Venus}/M_{Mercury}=$ 14.75), large iron-rich core \citep[$\sim$70-80$\%$ of its total mass:][]{hauck13} and dynamical isolation from Venus ($P_{Venus}/P_{Mercury}=$ 2.55) have been interpreted to imply that it formed in a different manner than the other terrestrial worlds.  Historically, Mercury analogs simultaneously possessing each of these qualities are rare to non-existent in N-body studies of the giant impact phase \citep{ray18_rev}.  While it is possible that Mercury's peculiar composition is the result of an energetic, mantle-stripping collision \citep{benz88,benz07,asphaug14}, the various proposed collisional scenarios \citep{jackson18,chau18} are highly improbable from a dynamical standpoint \citep{clement19_merc,fang20}.  Alternatively, several authors have explored the possibility that Mercury formed directly from an interior component \citep{chambers01,obrien06,lykawka17,lykawka19} of iron-enriched material \citep[e.g.:][]{ebel11,wurm2013}.  While adequate Mercury-Venus systems can be produced occasionally in such a scenario \citep{clement21_merc2}, the simulations of our current investigation are unable to capture either possibility by virtue of not incorporating a fragmentation model \citep[e.g.:][]{chambers13} and truncating the inner terrestrial disk outside of Mercury's modern semi-major axis.  Nevertheless, our 356 instability simulations yield a total of 14 Mercury-Venus analog systems as defined by \citet[][form exactly two planets inside of 0.85 au with $M_{Venus}/M_{Mercury}>$ 5.0 and $P_{Venus}/P_{Mercury}>$ 1.75]{clement21_merc2}.  Though adequate versions of Mercury are clearly rare, these results represent a marked improvement from the instability models of Paper I and Paper II ($\lesssim$ 1$\%$ of systems successful in this manner).  Moreover, the orbits of the majority of these Mercury-like planets are remarkably similar to that of the real planet.  Specifically, the median Mercury-analog eccentricity is 0.15 ($e_{Mercury}=$ 0.21) and the median inclination is 9.0$\degr$ ($i_{Mercury}=$ 7.0$\degr$); consistent with studies of the instability's effect on the circular orbit of an initially fully formed version of Mercury \citep{roig16}. 

While a complete analysis of the increased efficiency of forming Mercury in our new instability models is beyond the scope of our current investigation, it is clear that the primordial 2:1 Jupiter-Saturn resonance plays an important role in the process.  Specifically, 12 of the 14 Mercury analogs mentioned above are derived from simulations invoking the 2:1 instability model of \citet{clement21_instb}.  We speculate that a possible explanation for this result is interactions between the $\nu_{5}$ secular resonance and embryos in the Earth and Venus-forming regions during the instability.  Indeed, we expect perturbations from Jupiter's eccentricity in the inner solar system via $\nu_{5}$ to be significant throughout the entire duration of our 2:1 simulations as $e_{J}$ is initialized close to its modern value \citep[note that the giant planets eccentricities are likely higher in the gas disk phase if they are locked in the 2:1 MMR:][]{pierens14}.  In this manner, 10 of our Mercury analogs begin as embryos in the Earth-Venus forming region ($a\lesssim$ 1.0 au) with 0.05 $\lesssim M \lesssim$ 0.10 $M_{\oplus}$ before being liberated from the disk and stranded near Mercury's modern orbit by the instability.  Future numerical studies of the early instability scenario must utilize controlled instability evolutions \citep[e.g.:][]{nesvorny13} to thoroughly understand the viability of this Mercury-genesis scenario.

\begin{figure}
	\centering
	\includegraphics[width=.5\textwidth]{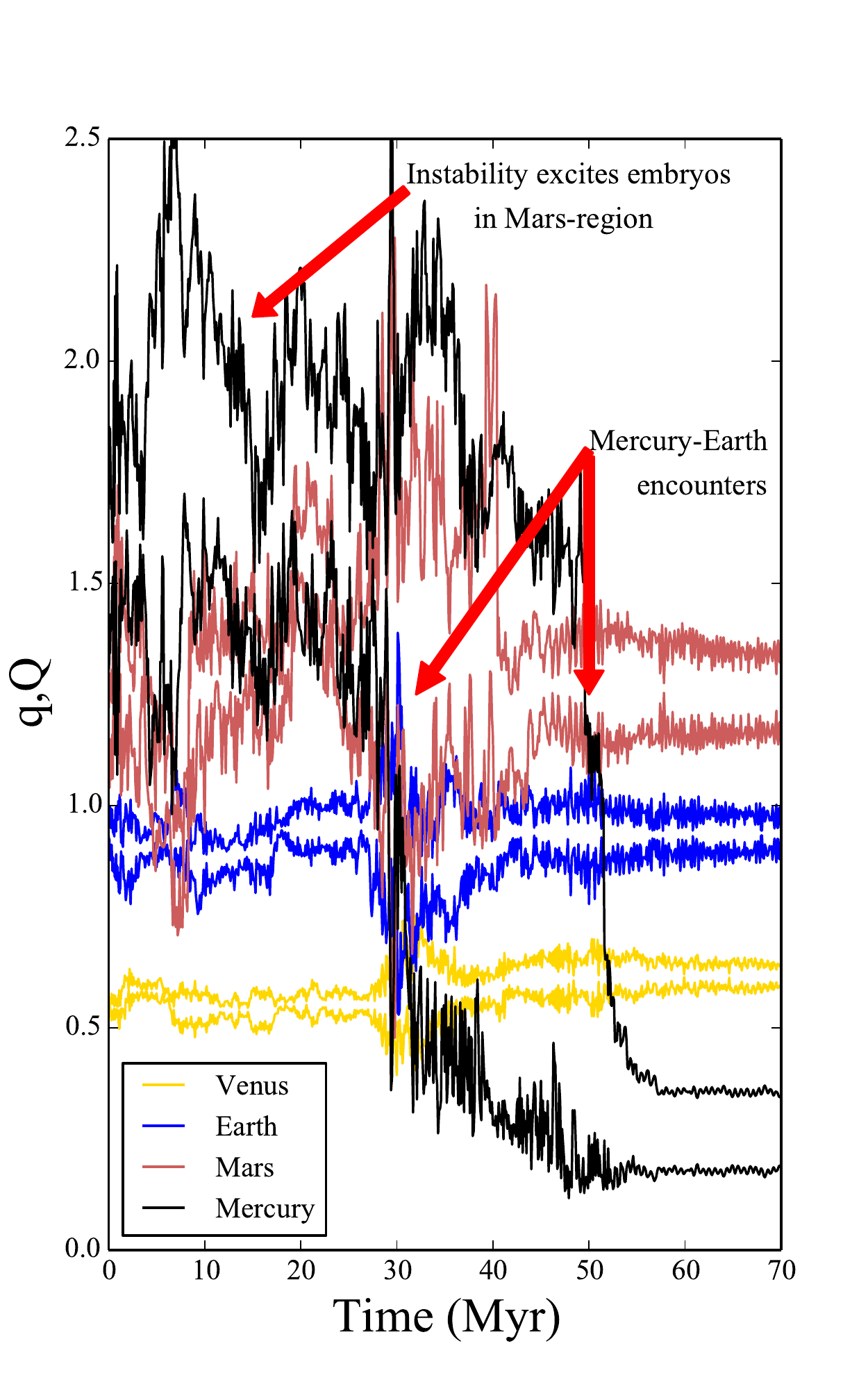}
	\caption{Example simulation from our C20/2:1/0Myr batch where an embryo in Mars region (black lines) is excited by the instability, and subsequently implanted inside of Venus' orbit as a Mercury analog through a series of close-encounters with the system's Earth analog.  The perihelia and aphelia for each of the four terrestrial planets formed in the simulation are plotted with respect to time.  The final terrestrial planet masses are $M_{Mercury}=$ 0.080 ($M_{Mercury,SS}=$ 0.055), $M_{Venus}=$ 0.99 ($M_{Venus,SS}=$ 0.815), $M_{Earth}=$ 0.69 and $M_{Mars}=$ 0.33 ($M_{Mars,SS}=$ 0.107) $M_{\oplus}$.  The simulation finishes with four giant planets and satisfies three of the four success criteria for the giant planets ($e_{55}=$ 0.018, $P_{S}/P_{J}=$ 2.48; compared with $e_{55,ss}=$ 0.044, $P_{S}/P_{J,ss}=$ 2.49)  described in \citet{nesvorny12} and \citet{clement21_instb}.}
	\label{fig:merc}
\end{figure}

Intriguingly, four of the 14 Mercury analogs formed in our instability models initially possess semi-major axes between 1.5 and $\sim$2.0 au.  An example of such a system is plotted in figure \ref{fig:merc}.  Mercury begins the simulation with $a=$ 1.57 au and $M=$ 0.040 $M_{\oplus}$.  After the eccentricities of all embryos in the Mars-region are excited by the instability (see also section \ref{sect:mars} and Paper I), a series of fortuitous close encounters with the growing Earth analog at $t\simeq$ 30 and 55 Myr conspire to scatter, and subsequently circularize Mercury's orbit around $a=$ 0.30 au (it should be noted that our integration time-step is insufficient to fully capture these dynamics).  While such a distant formation location for Mercury would likely imply a bulk composition similar to Mars', Mercury's large-scale chemical make-up remains somewhat unconstrained \citep{nittler17}.  Moreover, it is possible that Mercury's bulk composition was subsequently altered by a giant impact after its implantation.

\section{Discussion and Conclusions}

In this paper we revisited the early instability scenario for terrestrial planet formation developed in \citet{clement18} and \citet[][Papers I and II, respectively]{clement18_frag} with new models incorporating more accurate terrestrial disk initial conditions.  While the simulations of Paper I and Paper II exclusively considered an instability scenario where Jupiter and Saturn are initialized in the 3:2 MMR \citep{morby07,nesvorny11,nesvorny12}, our new models investigate both the 2:1 and 3:2 resonances \citep{pierens14,clement21_instb}.  Additionally, we develop a new computational pipeline that drastically increases the sample of fully evolved systems with Jupiter and Saturn-like orbits.  Through this process, our new models enable us to more accurately scrutinize the relationship between depletion in the terrestrial region and the fate of the outer solar system.

Our updated terrestrial disk models \citep[derived from high-resolution simulations of runaway growth in:][]{walsh19,clement20_mnras} are distinguished from those of Paper I and Paper II by the advanced evolutionary states of the proto-planets in the Earth and Venus-forming regions.  While a natural consequence of this is the rapid formation of the more massive terrestrial planets after the instability transpires, a reasonable fraction of our simulations satisfy constraints related to the inferred delayed occurrence of the Moon-forming impact \citep{kleine09,rudge10}.  Moreover, the formation of many of our Earth analogs culminate in a final giant impact involving roughly equal-mass impactors as proposed in \citet{canup12}.  Similarly, our contemporary models provide markedly improved matches to Mars' inferred growth timescale \citep{Dauphas11,kruijer17_mars} and the asteroid belt's highly depleted state.  Unfortunately, these models systematically struggle to replicate the low orbital eccentricities of Earth and Venus, and we highlight this shortcoming as one of the largest outstanding problems for terrestrial planet formation models to resolve \citep[see also:][]{nesvorny21_tp}. 

None of our simulations that adequately excites Jupiter's eccentricity finish with a large Mars analog ($M_{Mars} \gtrsim$ 0.3 $M_{\oplus}$).  Thus, our improved sample of solar system-like giant planet evolutions elucidate a strong correlation between Mars' final mass and the ultimate value of Jupiter's eccentricity.  Specifically, this represents, perhaps, the chief advantage of the early instability scenario over other proposed explanations for Mars' small mass \citep[e.g.:][]{walsh11,levison15,izidoro15,ray17sci}.  All terrestrial planet formation models must reconcile the inner solar system's post-accretion architecture and its modern structure within the Nice Model framework.  Thus, the fact that the large-scale mass distribution of the inner solar system is a common result in our models that adequately replicate the outer solar system is a compelling argument in favor of utilizing the instability to stunt Mars' growth.  These results also serve to further validate the primordial 2:1 Jupiter-Saturn resonance as a viable evolutionary pathway for the solar system.  

In spite of the substantial improvements in our contemporary study, several necessary simplifications limit the robustness of our conclusions.  First, to save compute time and enable a large statistical study, we generate initial planetesimal populations by interpolating from the distributions generated in \citet{walsh19} and \citet{clement20_psj}.  These disks are then input into independently-derived instability models; essentially applying a step-change to the outer planets' masses, semi-major axes and (in the case of our 2:1 models) eccentricities.  Moreover, our simulations do not account for collisional fragmentation \citep[e.g.:][]{chambers13}, which potentially affected Earth and Venus' eccentricities and Mercury's composition.  While some of our 2:1 models yield remarkably successful Mercury analogs (we plan to explore this result further in future work), our integrations are not properly tuned to fully resolve the innermost planets' formation.  Finally, certain geochemical constraints \citep[e.g.:][]{marty17} have been interpreted to disfavor the Nice Model's occurrence prior to the culmination of Earth's formation.  Prospective investigations of the early instability scenario should continue to improve the resolution of simulations in terms of the connection between supposed initial conditions and  disk models with an eye towards geochemical constraints and Mercury's formation.

\section*{Acknowledgments}

The authors thank two anonymous reviewers for insightful reports that improved the presentation of the results in this manuscript.  N.A.K. thanks the National Science Foundation for support under award AST-1615975 and NSF CAREER award 1846388.  S.N.R. acknowledges support from the CNRS’s PNP program and NASA Astrobiology Institute's Virtual Planetary Laboratory Lead Team, funded via the NASA Astrobiology Institute under solicitation NNH12ZDA002C and cooperative agreement no. NNA13AA93A.  Some of the computing for this project was performed at the OU Supercomputing Center for Education and Research (OSCER) at the University of Oklahoma (OU).  The authors acknowledge the Texas Advanced Computing Center (TACC) at The University of Texas at Austin for providing {HPC, visualization, database, or grid} resources that have contributed to the research results reported within this paper. URL: http://www.tacc.utexas.edu.  This work used the Extreme Science and Engineering Discovery Environment (XSEDE), which is supported by National Science Foundation grant number ACI-1548562. Specifically, it used the Comet system at the San Diego Supercomputing Center (SDSC) and the Bridges system, which is supported by NSF award number ACI-1445606, at the Pittsburgh Supercomputing Center \citep[PSC:][]{xsede}. Additional computation for the work described in this paper was supported by Carnegie Science's Scientific Computing Committee for High-Performance Computing (hpc.carnegiescience.edu).
\bibliographystyle{apj}
\newcommand{\sci}{$Science$ }
\bibliography{early_inst3.bib}
\end{document}